\documentclass[pdftex,12pt,authoryear]{elsarticle}

\usepackage[a4paper]{geometry}
\usepackage{hyperref}
\usepackage{color}
\usepackage{graphics}
\usepackage{xcolor}

\hypersetup{colorlinks, bookmarksopen,linkcolor = blue, citecolor = blue}

\newcommand{\rev}[1]{#1}

\usepackage{stmaryrd}
\usepackage{amssymb}

\newcommand\de[1]{\,{\mathrm d}#1} 
\newcommand{\set}[1]{{\mathbb #1}}

\newcommand{\eps}{\varepsilon}
\newcommand{\dmn}{\Omega}
\newcommand{\dmg}{\omega}
\newcommand{\bnd}{\Gamma}
\newcommand{\disp}{u}
\newcommand{\dispD}{u_{\rm D}}

\newcommand{\tdisp}{\widehat{\disp}}
\newcommand{\tdmg}{\widehat{\dmg}}

\newcommand{\vdisp}{\delta{\disp}}
\newcommand{\vdmg}{\delta{\dmg}}

\newcommand{\EF}{\mathcal{E}}
\newcommand{\DD}{\mathcal{D}}

\newcommand{\KV}{\set{K}}
\newcommand{\IV}{\set{Z}}

\newcommand{\gf}{g_{f0}} 
\newcommand{\Gf}{G_{f}} 
\newcommand{\gfn}{g_{f0}} 
\newcommand{\gff}{g_{f\infty}} 
\newcommand{\il}{\ell_0^2}
\newcommand{\ello}{\ell_{\omega}}

\newcommand{\dt}{\Delta t}
\newcommand{\rate}[1]{\dot{#1}}

\newcommand{\ft}{f_{\rm t}}

\newcommand{\WII}{W^{1,2}(\dmn)}

\newcommand{\half}{\mbox{$\frac{1}{2}$}}

\newcommand{\eqref}[1]{Eq.~(\ref{eq:#1})}
\newcommand{\figref}[1]{Figure~\ref{fig:#1}}

\newcommand{\Diss}{D}
\newcommand{\DissD}{Y}
\newcommand{\EE}{\Psi}

\newcommand{\beq}{\begin{equation}}
\newcommand{\eeq}{\end{equation}}
\newcommand{\bea}{\begin{eqnarray}}
\newcommand{\eea}{\end{eqnarray}}

\newcommand{\dx}{{\rm d}x} 

\newcommand{\jmp}[2]{\left\llbracket #1 \right\rrbracket_{#2}}
\newcommand{\strain}{\varepsilon}
\newcommand{\tstrain}{\widehat{\strain}}
\newcommand{\dd}{\mathrm{d}}

\newcommand{\secref}[1]{Section~\ref{sec:#1}}

\begin{document}

\begin{frontmatter}

\journal{arxiv}

\title{Localization Study of a Regularized \rev{Variational} Damage
Model}

\author[ctu]{Milan~Jir\'{a}sek}
\address[ctu]{%
Department of Mechanics, Faculty of Civil Engineering, Czech Technical
University in Prague,\\
Th\'{a}kurova 7, 166 29 Prague 6, Czech Republic}
\ead{Milan.Jirasek@fsv.cvut.cz}
\ead[url]{http://mech.fsv.cvut.cz/~milan}

\author[ctu]{Jan Zeman}
\ead{zemanj@cml.fsv.cvut.cz}
\ead[url]{http://mech.fsv.cvut.cz/~zemanj}

\begin{abstract}
The paper presents a detailed analysis and extended formulation of a
rate-independent regularized damage model proposed by~\cite{Mielke:2006:RID}.
Localization properties are studied in the context of a simple one-dimensional
problem, but the results reveal the fundamental features of the basic model and
of its modified versions. The initial bifurcation from a uniform solution is
described analytically while the complete failure process is studied
numerically. Modifications of the regularizing term and of the dissipation
distance are introduced and their effect on the global response is investigated.
It is shown that, with a proper combination of model parameters, a realistic
shape of the load-displacement diagram can be achieved and pathological effects
such as extremely brittle response or expansion of the damage zone accompanied
by stress locking can be eliminated.\end{abstract}

\end{frontmatter}

\paragraph{Keywords} 
damage, failure, dissipation, non-locality, \rev{variational approach}, bifurcation

\section{Introduction}

In engineering mechanics, \emph{damage} is understood as a load-induced
evolution of microstructural defects, resulting in a reduced macroscopic
material integrity. Phenomenological constitutive models of damage
\rev{characterize} such irreversible phenomena by \rev{an \emph{internal}
damage variable \citep{Kachanov:1958:TRP}, which is closely related to the
reduction of the secant modulus of elasticity}. Since the seminal contribution
of~\cite{Bazant:1976:IDSE}, it has been well-understood that such a description
within the framework of local (i.e. scale-free) continuum mechanics leads to an
ill-posed problem, resulting in localization of damage growth into an
arbitrarily small region. As a remedy to this pathology, a plethora of non-local
rate-independent continuum theories, based on integral, explicit and implicit
gradient approaches, have been proposed to introduce an \emph{internal
  length scale} into the description, see
e.g.~\cite{Bazant:2002:NIF} for a representative overview. Despite a significant
increase in objectivity offered by the enhanced continuum theories, the
non-local damage formulations often suffer from the fact that the non-local
variables are introduced into the model in an ad-hoc fashion, thus violating
basic constraints of thermodynamics. In addition, since the principle of local
action is no longer valid, such inconsistencies are rather difficult to detect,
especially in the multi-dimensional setting, e.g.~\cite{Simone:2004:IIP}.
Fortunately, as demonstrated by~\cite{Jirasek:1998:NMDF} and confirmed by a
number of independent studies,
e.g.~\citep{Peerlings:2001:CCNG,Jirasek:2003:DIT,DiLuzio:2005:SAL,Engelen:2006:EHO,Jirasek:2009:LPSSGPM1,Jirasek:2009:LPSSGPM2},
a simple one-dimensional study of the localization behavior can serve as a
convenient ``filter'' test, allowing to pinpoint various inconsistencies in the
formulation of a constitutive model. The same point of view has recently been
adopted by~\cite{Pham:2011:IUS} and~\cite{Pham:2013:FOD}, who investigated
various aspects of the response a wide class of energy-based gradient damage
models under displacement-controlled uniaxial tension. These works build on a
variational framework for local and gradient-based models developed
by~\cite{Pham:2010:TADI,Pham:2010:TADII}, in which evolution follows from
physically sound principles of stability, energy balance, and irreversibility,
expressed using a single energy functional. In particular, \cite{Pham:2011:IUS}
concentrates on the stability of homogeneous solutions, while in the follow-up
work~\citep{Pham:2013:FOD} the authors study in detail the behavior inside the
damaged zone and its implications for the structural response. In both cases,
the material constitutive law is incorporated in the model indirectly by means
of parametrized energy families with parameters adjusted to reproduce the local
stress-strain response of the material under investigation. The purpose of our
paper is to complement these developments with detailed localization studies for
gradient damage models based on the commonly used local stress-strain diagrams.
To this purpose, we start from the discussion of an elementary elastic-brittle
model \rev{regularized by the gradient of damage in the spirit
of~\cite{Fremond:1996:DGD};} see \secref{elastic-brittle}. Our description
builds on a general framework established by Mielke and co-workers, see
e.g.~\cite{Mielke:2005:ERI} for an overview, developed to study the evolution of
general irreversible rate-independent systems, which has been applied to
rigorous analysis of gradient damage models and their numerical approximation
\citep{Mielke:2006:RID,Bouchitte:2007:CDP,Mielke:2007:CDIV,Thomas:2010:DNE,Mielke:2011:CDE}.
The variational formulation presented in \secref{elastic-brittle} is thus based
on a stored energy functional, quantifying the reversibly stored energy, and a
dissipation distance accounting for the irreversible changes. The stored energy
is further decomposed into a standard~(elastic) part and a regularizing part
which introduces a characteristic length into the formulation.

\secref{sec3}
presents a study of the localization behavior of the model, utilizing arguments
of \rev{local} incremental energy minimization. Following our recent developments~\citep{Jirasek:2013:LAV},
in \secref{elastic-brittle-regularity} we show that the damage profiles during
the damage evolution must be continuously differentiable in space, thereby
justifying the assumption made by~\cite[Remark~2]{Pham:2013:FOD}, and derive the
continuity conditions at the interface between elastic and damaging regions, as
well the governing equations to be satisfied in the region experiencing damage.
These conditions are employed in \secref{elastic-brittle-onset} to characterize
the elastic response, in \secref{elastic-brittle-governing} to obtain an
analytical solution to the damage profile at the onset of damage, and in
\secref{elastic-brittle-numerical} to study the response at later stages  by
means of a numerical procedure described in \ref{app:algorithm}. It turns out
that the model is regularized in the sense that the energy dissipation is
finite, but the global response is extremely brittle, especially at late stages
of the failure process. This motivates the search for modifications which could
lead to load-displacement diagrams that better correspond to the actual behavior
of quasibrittle materials.

In Section~\ref{sec4}, the elastic-brittle core of the
model is replaced by linear or exponential softening via modifications of
the  dissipation distance. In Section~\ref{sec5}, an elastic-brittle model 
with the regularizing part of the stored energy dependent on the gradient of
a modified internal variable is developed and its alternative interpretation
in terms of a  variable characteristic length is suggested. 
Finally, Section~\ref{sec6} combines the linear or exponential softening 
with variable characteristic length. 

\section{\rev{Variational} formulation of elastic-brittle
model}\label{sec:elastic-brittle}

\begin{figure}[t]
\includegraphics[scale=.925]{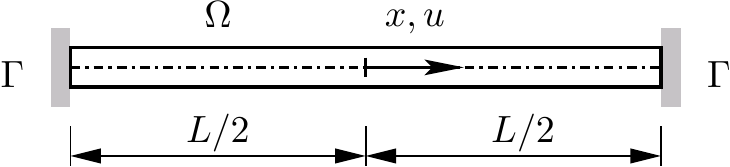}
\hfill
\includegraphics[scale=.925]{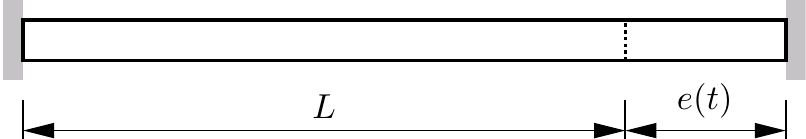}
\caption{Bar under uniaxial displacement-controlled tension.}
\label{fig:uni_bar}
\end{figure}

We consider a prismatic bar of initial length $L$, subjected to
displacement-controlled uniaxial tensile loading. In the sequel, the bar will be
represented by the interval $\dmn = (-L/2;L/2)$, with boundary
$\bnd=\{-L/2,L/2\}$ (consisting of two points) subjected to the Dirichlet
loading $\dispD(t) : \Gamma \rightarrow \set{R}$, where $t\in [ 0; T ]$ denotes
the (pseudo-) time; see~\figref{uni_bar}. For the sake of simplicity, we denote
by $e$ the bar elongation (change of length), i.e., we set $e(t) =
\dispD(t,L/2)-\dispD(t,-L/2)$ in what follows.

Following the standard thermodynamic approach to constitutive
modeling, summarized e.g.\ in Chapter~25 of \cite{Jirasek:2002:IAS}, a state of the
system is described using \emph{admissible} displacement and damage
fields $\tdisp : \dmn \rightarrow \set{R}$ and $\tdmg : \dmn
\rightarrow \set{R}$. Formally, we write
\begin{eqnarray}
\tdisp \in \KV(t) & = & \left\{ \tdisp \in \WII, \tdisp(x)|_\bnd = \dispD( t )
\right \} \\ \tdmg  \in \IV & = &
\left\{ \tdmg \in \WII, 0 \leq \tdmg(x) \leq 1 \mbox{ in } \dmn \right\}
\end{eqnarray}
where $\KV(t)$ denotes the set of kinematically admissible displacements at
time $t$, $\IV$~stands for the set of admissible damage fields, and $\WII$ is
the Sobolev space of functions with square-integrable \rev{distributional}
derivatives; see e.g.~\cite{Rektorys:1982:MDT}. 

Within the adopted \rev{variational}
framework~\citep{Mielke:2006:RID}, the constitutive description of the damage model is
based on
\begin{enumerate}
\item
the \emph{stored energy} functional
\begin{equation}\label{eq:GE}
\EF(\tdisp, \tdmg ) 
=\EF_{\rm std}(\tdisp, \tdmg )+\EF_{\rm reg}(\tdmg )
\end{equation}
with the \emph{standard} part~$\EF_{\rm std} : W^{1,2}(\dmn) \times \IV
\rightarrow \set{R}$ and the \emph{regularizing} part $\EF_{\rm reg} : \IV \rightarrow
\set{R}$ respectively defined as
\begin{eqnarray}
\EF_{\rm std}(\tdisp, \tdmg ) &=&\half \int_\dmn (1 - \tdmg(x) ) E  \tdisp'^2(x) \de x
\label{eq:GEstd}
\\
\EF_{\rm reg}(\tdmg )&=&\half \int_\dmn \gf \il \tdmg'^2(x) \de x
\label{eq:GEreg}
\end{eqnarray}
where $\tdisp'$ corresponds to an admissible strain field $\tstrain$, 
\item
the \emph{dissipation distance} $\DD : \IV \times \IV \rightarrow \set{R}\cup 
\{ +\infty \}$
\begin{equation}\label{eq:GDD}
\DD( \tdmg_1, \tdmg_2 )
=
\left\{ 
\begin{array}{ll}
\displaystyle \int_\dmn \gf \left( \tdmg_2(x) - \tdmg_1(x) \right) \de x
  & \mbox{if } \tdmg_2 \geq \tdmg_1 \mbox{ in }
  \dmn \\ +\infty & \mbox{otherwise}
\end{array}
\right.
\end{equation}
\end{enumerate}

Physically, $\EF$ represents the energy reversibly stored in the system
and $\DD$ is the energy dissipated by changing the damage field from $\tdmg_1$ to
$\tdmg_2$. The reversibly stored energy consists of the standard part $\EF_{\rm
std}$ and the regularizing part $\EF_{\rm reg}$; the latter depends on the damage
gradient and acts as a localization limiter. Note that $\EF_{\rm reg}$  vanishes
for uniform damage states. In Eqs.~(\ref{eq:GEstd})--(\ref{eq:GDD}),
$E$~[Pa] denotes the Young modulus, $\gf$~[Jm$^{-3}$] is the
amount of energy needed to disintegrate a unit volume of the material, and
$\ell_0$~[m] is a characteristic material length, which reflects the size
and spacing of dominant heterogeneities in the microstructure. Later it will
become clear that the ``$+\infty$'' term appearing in~\eqref{GDD} enforces
irreversibility of damage evolution, i.e., ensures that the damage variable
cannot decrease in time.

Now, given the Dirichlet loading $\dispD$, functionals $\EF$ and
$\DD$ and initial data $\bar{\disp}_0 \in \KV(0)$ and $\bar{\dmg}_0 \in \IV$, the
\emph{energetic solution} of the damage problem is provided by
time-dependent fields $\disp(t) \in \KV(t)$ and $\dmg(t) \in \IV$
satisfying~\citep{Mielke:2005:ERI}:

\begin{description}

\item[Global stability:] for all $t \in [0; T]$, $\tdisp \in \KV(t)$
  and $\tdmg \in \IV$
\begin{equation}\label{eq:GS}
\EF(\disp(t), \dmg(t) ) 
\leq
\EF(\tdisp, \tdmg ) 
+
\DD( \dmg(t), \tdmg )
\end{equation}

\item[Energy equality:] for all $t \in [0; T]$
\begin{equation}\label{eq:EE}
\EF(\disp(t), \dmg(t)) 
+ 
\mathrm{Var}_\DD(\dmg, [0;t])
=
\EF(\disp(0), \dmg(0)) 
+ 
\int_0^t 
\int_{\Gamma}R(s)\dot{\disp}_\mathrm{D}(s)\de\Gamma
\de s
\end{equation}
where 
$$
\mathrm{Var}_\DD(\dmg, [0;t])
=
\sup
\sum_{j=1}^J \DD(\dmg(t_{j-1}),\dmg(t_j))
$$
is the energy dissipated during the time
interval~$[0;t]$
(with the supremum taken over all partitions of $[0;t]$ in the form
$0=t_0<t_1<...<t_{J-1}<t_J=t$), $R$ are the reactions arising at the
boundary, and the product $R\,\dot{\disp}_\mathrm{D}$ is the external power.

\item[Initial conditions:]  
\begin{eqnarray}\label{eq:IC}
\disp(0) = \bar{u}_0 & \mathrm{and} & \dmg(0) = \bar{\omega}_0
\end{eqnarray}
For simplicity, we will consider initial data $\bar{u}_0\rev{=u_D(0)}=0$ and
$\bar{\omega}_0=0$, which correspond to an initial undeformed and damage-free state. Note that,
for initial conditions in such a form, the first term on the right-hand 
side of \eqref{EE} vanishes and can be dropped. 
\end{description}

Although conditions~(\ref{eq:GS})--(\ref{eq:IC}) present a formal
definition of the energetic solution, the actual analysis will be performed
using the time discretization technique, see e.g.~\cite{Rektorys:1982:MDT} for a
nice exposition. To that end, we consider a partition of the time
interval $[0;T]$
\begin{equation}\label{eq:partioning}
0 = t_0 < t_1 = t_0 + \dt_1 < \ldots < t_{N-1} + \dt_{N}
= t_{N} = T
\end{equation}
and recursively solve the minimization problem
\begin{equation}\label{eq:incr_min}
\left( \disp_k, \dmg_k \right)
\in 
\mbox{Arg} \min_{(\tdisp, \tdmg) \in \KV(t_k) \times \IV}
\left[
\EF(\tdisp, \tdmg ) 
+
\DD( \dmg_{k-1}, \tdmg )
\right]
\mbox{ for }k = 1, 2, \ldots, N
\end{equation}
Here, e.g. $\disp_k$ and $\dmg_k$ abbreviate $\disp(t_k)$ and $\dmg(t_k)$, and
$\mbox{Arg} \min$ denotes the set of (possibly non-unique) minimizers of the
incremental problem. 

\rev{The main asset of the energetic solution concept is that, under
reasonable data qualification, the solutions of the time-discretized problem
converge to the energy-conserving energetic solution as $\max\limits_k \dt_k
\rightarrow 0$,
see~\cite{Mielke:2006:RID,Bouchitte:2007:CDP,Mielke:2007:CDIV,Thomas:2010:DNE,Mielke:2011:CDE}.
This convergence, however, requires the minimization in the stability
condition~(\ref{eq:GS}) and the incremental problem~(\ref{eq:incr_min}) to be
performed \emph{globally}, implying that damage may initiate and propagate
\emph{before} the energy threshold $g_{f0}$ is exceeded at any point of the
structure---a phenomenon that is often found difficult to justify from the
physical point of view. More physical, but also more technically involved,
solution concepts thus rely on a suitable \emph{local} energy minimization, see
e.g.\ the overviews~\citep{Mielke:2011:DEM,Braides:2014:LM} for the treatment of general
rate-independent systems and~\citep{Roubicek:2015:MDL,Vodicka:2014:EMD} for
damage- and delamination-related studies.

The approach we adopt here follows the earlier work by~\cite{Pham:2013:FOD} and
consists of two adjustments to the introduced setting. First, we search for
\emph{critical points} of the incremental problem, corresponding to the
so-called first-order stability conditions according to the terminology
introduced in~\cite[Section~2]{Pham:2013:FOD}. Second, once the threshold
$g_{f0}$ is reached for at least one point within the structure, the evolution is driven by
the maximum of \emph{damage variable} $\omega$; the Dirichlet data $u_D$ are then
obtained to be compatible with the resulting damage profiles,
see~\cite[Section~4]{Pham:2013:FOD}. As a result, we obtain solutions that are
smooth in time and automatically satisfy the energy equality~(\ref{eq:EE}),
\citep[Property~1]{Pham:2013:FOD}. Note that issues of local stability of the
localized solutions and the relation between the present results and the
alternative solution concepts are out of the scope of this paper and will be
investigated separately.}

\section{Analysis of localization behavior}\label{sec:sec3}

Having introduced the essentials of the \rev{variational} framework,
we will now proceed towards the main goal of this contribution -- localization analysis of a simple
uniaxial tensile test based on \rev{a critical point of} the incremental
variational principle~(\ref{eq:incr_min}), formulated for the time interval
$[t_{k-1}; t_k]$. 

\subsection{Regularity study}\label{sec:elastic-brittle-regularity}

The necessary optimality conditions of
problem~(\ref{eq:incr_min}) read
\begin{eqnarray}\label{eq:mj3}
\int_\dmn
\left[
\vdisp'(x) ( 1 - \dmg_k(x) ) E \disp_k'(x)
-
\half
\vdmg(x) E \disp_k'^2(x) 
\right.
\nonumber \\ 
+
\left.
\vdmg'(x) \gf \il \dmg_k'(x)
+
\vdmg(x) \gf 
\right]
\de x 
\geq
0
\end{eqnarray}
for all admissible displacement variations $\vdisp\in\WII$ vanishing on
$\bnd$ and for arbitrary damage variations $\vdmg\in\WII$ satisfying 
\begin{equation}\label{eq:omega_adm}
\dmg_{k-1}(x) \leq \dmg_k(x) + \vdmg(x) \leq 1
\mbox{ for } 
x \in \dmn
\end{equation}
After integration by parts, we formally obtain\footnote{%
Here, we need to assume that the solutions
$\disp$ and $\dmg$ are piecewise $C^2$ continuous in $\dmn$.
Also note that $\disp$, $\dmg$, $\vdisp$, and $\vdmg$ are continuous
functions on $\dmn$, by virtue of the embedding $W^{1,2}(\dmn) \subset
C^0(\rev{\overline{\dmn}})$.}
\begin{eqnarray}\label{eq:mj4}
-
\int_\dmn
\vdisp(x)
\Bigl( ( 1 - \dmg_k(x) ) E \disp_k'(x) \Bigr)'
\de x
-
\int_\dmn
\vdmg(x)
\left(
\half
 E \disp_k'^2(x) 
+
\gf \il \dmg_k''(x)
-
\gf 
\right)
\de x 
\nonumber
\\
-
\sum_{i}\jmp{%
\vdisp(x)
\bigl( ( 1 - \dmg_k(x) ) E \disp_k'(x) \bigr)
}{x_i}
-
\sum_{j}\jmp{%
\vdmg(x)
\gf \il
\dmg_k'(x)
}{x_j}
\nonumber
+  
\left[ \vdmg(x) \gf \il \dmg_k'(x) \right]_{\bnd}
\geq 
0
\\
\label{eq:optim_cond}
\end{eqnarray}
where 
\begin{equation}
\jmp{f(x)}{x_i}
=
\lim_{x \rightarrow x_i^+} f(x)
-
\lim_{x \rightarrow x_i^-} f(x)
\end{equation}
denotes the jump of function $f$ at its
discontinuity point $x_i$,
and $[ f(x) ]_\bnd = f(L/2) - f(-L/2)$. The sums in~\eqref{optim_cond} are
taken over all discontinuity points of arguments in the double brackets,
if such discontinuities exist.

The arbitrariness of variations $\vdisp$ and their continuity give rise
to conditions
\begin{eqnarray}
\Bigl( \left(1 - \dmg_k(x)\right) E \disp_k'(x) \Bigr)' =  0 \mbox{ for } x \in
\dmn, && 
\jmp{\left(1 - \dmg_k(x)\right) E u_k'(x)}{x_i} = 0
\end{eqnarray}
so that the stress, defined as
$
\sigma_k(x) 
= 
\left(1 - \dmg_k(x) \right) 
E \disp_k'(x)
$,
remains constant along the whole bar: 
\begin{equation}\label{eq:sigma_const}
\left(1 - \dmg_k(x) \right) 
E \disp_k'(x)
=\sigma_k=\mbox{ const.}
\end{equation}
Since the displacement derivative $\disp_k'(x)$ has the physical meaning
of strain, it will be occasionally denoted as $\eps_k(x)$.

It will be convenient for the forthcoming discussion to decompose the domain
$\dmn$ at each time $t_k$ into three disjoint sets\footnote{It is not necessary to consider the case 
when $\dmg_{k-1}(x)=\dmg_k(x)=1$. As soon as the damage reaches 1 at one point (cross section), 
no force can be transmitted by the bar and the failure process is complete.}
\begin{eqnarray}
\dmn_{e,k} 
& = &
\left\{ x \in \dmn : \omega_{k-1}(x) = \omega_k(x) < 1 \right\}
\\
\dmn_{d,k} 
& = &
\left\{ x \in \dmn : \omega_{k-1}(x) < \omega_k(x) < 1 \right\}
\\
\dmn_{f,k}
& = &
\left\{ x \in \dmn : \omega_{k-1}(x) <\omega_k(x) = 1 \right\}
\end{eqnarray}
corresponding to the elastic, damaging and fully damaged regions,
respectively. Note that the elastic region is understood as the region with a
vanishing damage {\em increment} between $t_{k-1}$ and $t_k$, not necessarily
with a vanishing  damage {\em value}. In order to meet the admissibility
condition~(\ref{eq:omega_adm}), the variation $\vdmg(x)$ can only be
non-negative for $x \in \dmn_{e,k}$, non-positive for $x \in \dmn_{f,k}$, and of
an arbitrary sign for $x \in \dmn_{d,k}$. Therefore, the Karush-Kuhn-Tucker optimality
conditions, e.g.~\cite[Section~5.2]{Jahn:2007:ITNO}, with respect to
$\vdmg$, implied by \eqref{optim_cond}, read
\begin{eqnarray}
\gf \il \dmg_k''(x) + \half E \disp_k'^2(x) \le \gf, &
\jmp{\dmg_k'(x)}{x_j} \le 0 &
\mbox{for } x, x_j \in \dmn_{e,k}
\label{eq:EL1}
\\
\gf \il \dmg_k''(x) + \half E \disp_k'^2(x) = \gf, &
\jmp{\dmg_k'(x)}{x_j} = 0 &
\mbox{for } x, x_j \in \dmn_{d,k}
\label{eq:EL3}
\\
\gf \il \dmg_k''(x) + \half E \disp_k'^2(x) \ge \gf, &
\jmp{\dmg_k'(x)}{x_j} \ge 0 &
\mbox{for } x, x_j \in \dmn_{f,k}
\label{eq:EL2}
\end{eqnarray}
In addition, at the boundary $\bnd=\{-L/2,L/2\}$ 
we obtain
\begin{eqnarray}\label{eq:EL5}
\dmg_k'(-L/2) = 0, && 
\dmg_k'(L/2) = 0 
\end{eqnarray}
If a boundary point  belongs to $\Omega_{d,k}$, the condition of vanishing derivative $\dmg_k'$
can be directly deduced from variational inequality (\ref{eq:mj4}), 
using the argument that the variation $\delta\dmg$ can have an arbitrary sign.
If, on the other hand, a boundary point  belongs to $\Omega_{e,k}$, then the variation $\delta\dmg$ 
at that point is nonnegative and (\ref{eq:mj4}) implies $\dmg_k'(-L/2)\leq 0$ at the left boundary,
or $\dmg_k'(L/2) \geq 0$ at the right boundary. But even then the derivative must vanish, as can be shown
by contradiction. Initially, damage is identically zero in the whole bar and the condition of zero derivative
at the boundary is satisfied. Suppose that $k$ is the first step in which  $\dmg_{k}'(-L/2)$ becomes negative. Since
the boundary point $-L/2$ must belong to $\Omega_{e,k}$ 
(otherwise the first condition in (\ref{eq:EL5}) would apply by virtue of (\ref{eq:mj4})),
the damage increment at that point must be zero and $\dmg_k(-L/2)=\dmg_{k-1}(-L/2)$. But then
 $\dmg_k'(-L/2)$ cannot be smaller than $\dmg_{k-1}'(-L/2)$ without violating the condition that
the damage increments must not be negative. Since $\dmg_{k-1}'(-L/2)=0$, we conclude that
 $\dmg_k'(-L/2)$ cannot be negative. Similar arguments (just with the opposite inequalities) can be
used at the right boundary. This proves that homogeneous Neumann boundary conditions (\ref{eq:EL5})
are universally applicable, independently of the state of the material at the boundary. 

\subsection{Elastic response}\label{sec:elastic-brittle-onset}

According to \eqref{IC} with the standard initial values $\bar{u}_0=0$ and $\bar{\omega}_0=0$, 
the loading program is assumed to start from an initial
undeformed and damage-free state, and so
$u_0(x)=0$, $\eps_0(x)=u_0'(x)=0$, $\dmg_0(x)=0$ and $\sigma_0=0$. The first condition
in (\ref{eq:EL1}) is then satisfied as a strict inequality and the entire bar is in
an elastic state, $\dmn_{e,0} = \dmn$. As the applied displacement at the
boundary is increased, the bar deforms and, up to a certain level of loading,
remains elastic. As long as the damage remains zero, \eqref{sigma_const}
implies that
\begin{equation}\label{eq:const_strain}
\eps_k(x) \equiv u_k'(x) =  \frac{\sigma_k}{E}
\end{equation}
where $\sigma_k$ is the current stress level. This means that, before the onset
of damage, the strain is uniform.
Such a solution remains admissible as long as the first part
of (\ref{eq:EL1}) is satisfied. In view of \eqref{const_strain}, this condition can
be rewritten as
\begin{equation}\label{eq:elastic_states}
\frac{\sigma_k^2}{2E}\le g_{f0}
\end{equation}
The stress level can be linked to the applied displacements at the
boundary, since the bar elongation (change of length) is given by
\begin{equation}
e_k 
= 
\int_\dmn
\eps_k(x)
\de x
=
\frac{\sigma_k}{E}
\int_\dmn
\de x
= 
\frac{\sigma_k L}{E}
\end{equation}
The onset of damage is attained when \eqref{elastic_states} is satisfied as an
equality, i.e., when
\begin{equation}
\sigma_k=\sqrt{2E\gf}\equiv \ft
\end{equation}
where $\ft$ denotes the tensile strength, derived from the given elastic modulus $E$ and
parameter $\gf$.
The corresponding bar elongation at the onset of damage is
\begin{equation}
e_{\rm od} = \frac{\ft L}{E}
\end{equation}
When the actual elongation exceeds this value, the bar cannot remain elastic along its entire length because
condition (\ref{eq:elastic_states}) would be violated. 

In what follows, the time corresponding to the onset of damage is
referred to as $t_{\rm od}$, and our objective is to characterize the response of the
bar for $t_k \geq t_{\rm od}$. 
In fact, since the elastic solution is described by very simple closed-form expressions,
the loading process from the initial state to the onset of damage can be handled by a single increment
and we can select time $t_1$ such that $e_1 = e_{\rm od}$.

\subsection{Damage evolution}\label{sec:elastic-brittle-governing}

When the damage threshold is exceeded, the strain field cannot remain uniform.
 According to \eqref{sigma_const}, uniform strain would imply uniform
 damage, $\dmg_k(x)=$ const., but then the first term on the left-hand side of
 \eqref{EL3} would vanish and the condition
would reduce to
 \begin{equation}
\half E\eps_k^2 = \gf
\end{equation}
which is satisfied exclusively for the state at the onset of damage.

In order to derive the governing equation for the damage field in
the damaging zone $\Omega_{d,k}$, we divide \eqref{EL3} by $\gf$ and, employing
identity~(\ref{eq:sigma_const}), rewrite it in the form of a non-linear ordinary
differential equation
\begin{equation}\label{eq:dmg_non_lin}
\il \dmg_k''(x) + \frac{\mu_k}{( 1 - \dmg_k(x) )^2} = 1
\end{equation}
where the newly introduced dimensionless parameter
\begin{equation}
\mu_k = \frac{\sigma_k^2}{2 E \gf} 
\end{equation}
provides a convenient parametrization to the damage evolution, as will be shown
next. At the onset of damage, $\dmg_1(x)=0$, $\sigma_1=\ft=\sqrt{2E\gf}$, and $\mu_1=\ft^2/2E\gf=1$.

\eqref{dmg_non_lin} is an ordinary second-order differential equation\rev{,
which should be satisfied in the unknown damaging zone $\Omega_{d,k}$, and as
such it represents a \emph{free-boundary problem},
e.g.~\citep{Friedman:1982:VPF}. Therefore, apart from the solution $\omega_k$
itself, we also search for the so-called \emph{noncoincidence set}
$\Omega_{d,k}$ and the continuity (or regularity) conditions at the
\emph{free-boundary} $\partial \Omega_{d,k}$ separating the damaging zone from
the elastic zone.} The advantage of the variational format of the model is that
the optimality conditions~(\ref{eq:EL1})--(\ref{eq:EL2}) encode the appropriate
conditions at the elastic-damaging interface.

In what follows, we assume that the damaging region is a single interval of
length $L_{d,k}$ and that it is shorter than the entire
bar (i.e., that $L_{d,k}<L$), to ensure that the physical boundary of the bar is located in the elastic
region.\footnote{Solutions localized at the boundary or in multiple non-overlapping
damaging intervals can be treated using this basic scenario, as
discussed in detail by~\cite{Pham:2013:FOD}.} \rev{However, for a perfectly
uniform bar (i.e., a bar with constant cross-sectional area and constant material
properties along the length) 
the solution would not be unique because the damage zone can be
arbitrarily translated along the bar. Physically, the actual position of the
damage zone would be determined by random imperfections. Without loss of
generality, we can thus assume that the damage zone}
\begin{equation}\label{eq:symmetric_damage}
\Omega_{d,k}
= ( -L_{d,k}/2;L_{d,k}/2 )
\end{equation}
\rev{is centered at the origin of
the spatial coordinates.}
Note that $\Omega_{d,k}$ is an open interval, and the points $\pm L_{d,k}/2$ belong to $\Omega_{e,k}$.

For the right boundary of the damaging zone $\Omega_{d,k}$ (i.e., for the point $L_{d,k}/2$, which itself
belongs to $\Omega_{e,k}$), we infer from the second part of (\ref{eq:EL1}) that
\begin{equation}
\lim_{x \rightarrow L_{d,k}^-/2} \dmg'_k(x)
\geq
\lim_{x \rightarrow L_{d,k}^+/2} \dmg'_k(x)
\label{eq:ineq}
\end{equation}
Let us first assume that the damage zone expands, i.e., that $L_{d,k}\ge L_{d,k-1} \ge\ldots \ge L_{d,2}$
(the case of a contracting damage zone will be treated later, in Section~\ref{sec:elastic-brittle-numerical}). 
In this case, $\dmg_k(x)=0$ for all $x\ge L_{d,k}/2$ and the right-hand side of (\ref{eq:ineq}) vanishes.
The condition then reduces to
\begin{equation}
\lim_{x \rightarrow L_{d,k}^-/2} \dmg'_k(x)\geq 0
\label{eq:ineq4}
\end{equation}
Since the value of damage at point $L_{d,k}/2$ is zero and the values to the left of this point are nonnegative,
the  limit on the left-hand side of (\ref{eq:ineq4}), which represents the derivative from the left,
cannot be positive. Consequently, $\dmg'_k$ must vanish at $x=L_{d,k}/2$, and by similar arguments we can show
that it must also vanish at $x=-L_{d,k}/2$,
which means that the solution preserves continuous differentiability. 

In summary, the solution must satisfy conditions
\bea\label{eq:ineq5}
\dmg_k(-L_{d,k}/2)=\dmg_k(L_{d,k}/2)= 0, \hskip 10mm
\dmg_k'(-L_{d,k}/2)=\dmg_k'(L_{d,k}/2)=0
\eea
These are the \rev{free-boundary conditions} to be
imposed on the solution of differential equation (\ref{eq:dmg_non_lin}). 
In general one could consider the coordinates of the left and right boundary
points of the damage zone as independent unknowns and conditions analogous to
(\ref{eq:ineq5}) would be used to determine these two unknowns plus two
integration constants present in the general solution of \eqref{dmg_non_lin}.
\rev{The assumed symmetry of
the damaging zone,~\eqref{symmetric_damage}, implies that one} of conditions
(\ref{eq:ineq5}) becomes redundant and the remaining three can be
used to determine the size of the damage zone, $L_{d,k}$, and two integration
constants. An interested reader is invited to consult~\cite{Jirasek:2013:LAV,Rokos:2015:LAE} for related results in the context of softening gradient plasticity with non-uniform data.
\eqref{dmg_non_lin} is nonlinear and cannot be solved in closed form.
However, if we consider the state just at the onset of damage and rewrite the equation in the rate
form, we obtain a linear equation that can be handled analytically.
To this end, let us assume that the damage evolution is sufficiently smooth\footnote{Later it will be shown that,
due to the
regularizing energy term~(\ref{eq:GEreg}), the response of the bar is continuous despite
the brittle character of the underlying non-regularized material model.}
in time,
such that we can introduce the damage rate
\begin{equation}
\rate{\dmg}_k(x)
=
\lim_{\Delta t_{k+1} \rightarrow 0^+}
\frac{\dmg(t_{k+1}) - \dmg(t_k)}{\Delta t_{k+1}},
\end{equation}
and rewrite \eqref{dmg_non_lin} in the rate
form as
\begin{equation}\label{eq:rate_form}
\il \rate{\dmg}_k''(x) 
+
2 \frac{\mu_k}{(1-\dmg_k(x))^3}\rate{\dmg}_k(x)
=
-
\frac{\rate{\mu}_k}{(1-\dmg_k(x))^2}
\end{equation}
This is an ordinary linear second-order differential equation for the damage
rate $\rate{\dmg}_k$ with possibly non-constant coefficients. Recall that 
damage irreversibility requires $\rate{\dmg}_k \geq 0$.

\begin{figure}[ht]
\centering
\begin{tabular}{cc}
(a) & (b) \\
\includegraphics{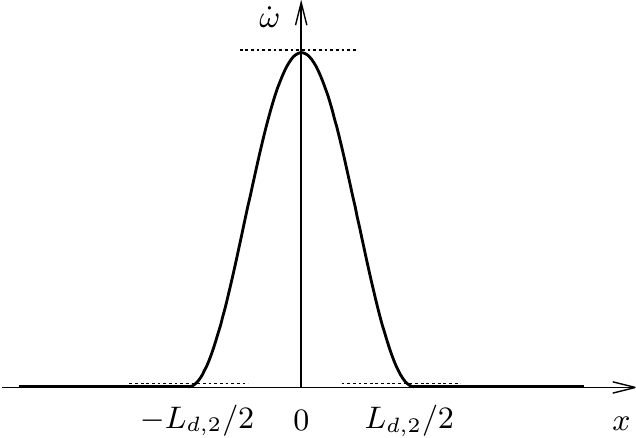} &
\includegraphics{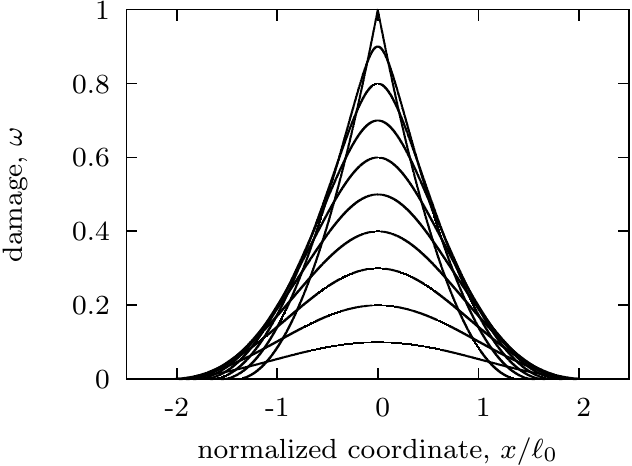}
\end{tabular}
\caption{(a) Sketch of the damage rate profile at the onset of damage; (b)~inadmissible evolution of
the damage profile computed using conditions  (\ref{eq:ineq5}) in terms of the total
damage ($\dmg=0$ and $\dmg'=0$ at the boundary of the damage zone).}
\label{fig:bif_shp}
\end{figure}

As already explained, at the onset of damage we have $\dmg_1(x) = 0$ and $\mu_1 =1$. 
Then, \eqref{rate_form} further simplifies to
\begin{equation}
\il \rate{\dmg}_1''(x) + 2 \rate{\dmg}_1(x) = - \rate{\mu}_1
\end{equation}
which has the general solution
\begin{equation}
\rate{\dmg}_1(x) 
= 
-\half \rate{\mu}_1
+ C_1 \cos \frac{\sqrt{2}x}{\ell_0}+ C_2 \sin \frac{\sqrt{2}x}{\ell_0}, \hskip 10mm x\in\Omega_{d,2}
\label{eq:gensol}
\end{equation}
Here, $C_1$ and $C_2$ denote the integration constants, which 
should be determined from appropriate boundary conditions.
Let us postulate these conditions by analogy with (\ref{eq:ineq5}) in the form
\begin{eqnarray}\label{bcondrate}
\rate{\dmg}_1(-L_{d,2}/2) = \rate{\dmg}_1(L_{d,2}/2) = 0, 
&&
\rate{\dmg}_1'(-L_{d,2}/2) = \rate{\dmg}_1'(L_{d,2}/2) = 0
\end{eqnarray}
Substituting the general solution (\ref{eq:gensol}) into (\ref{bcondrate}), we obtain integration constants
\beq
C_1 = -\half \rate{\mu}_1,
\hskip 10mm
C_2 = 0
\eeq
as well as the initial size of the damage zone
\beq
L_{d,2} = \sqrt{2} \pi \ell_0
\eeq
The rate of the damage variable at the onset of damage, plotted in \figref{bif_shp}(a), is therefore given by
\begin{eqnarray}\label{eq:onset_solution}
\rate{\dmg}_1(x) 
= 
-\half \rate{\mu}_1 
\left( 
  1 + \cos\frac{\sqrt{2}x}{\ell_0} 
\right) 
& \mbox{ for } &
x \in \dmn_{d,2}
\end{eqnarray}
This is consistent with the damage irreversibility constraint $\rate{\dmg}_1
\geq 0$ only if $\rate{\mu}_1 < 0$, which corresponds to a
negative stress rate (the case of $\rate{\mu}_1 = 0$ can be excluded because it 
can occur only if the state of the bar does not change at all and
all variables remain constant in time). Consequently, the onset of damage immediately results into a
\emph{softening} response. 

\subsection{Numerical solution}\label{sec:elastic-brittle-numerical}
%
\eqref{onset_solution} describes the initial damage rate at the onset of localization,
but the damage profiles at later stages of the failure process should be solved from \eqref{dmg_non_lin}. 
Since the analytical solution is not available, 
we resort to a numerical procedure using a variant of the shooting method, the details of
which are given in
\ref{app:algorithm}.

Let us recall that the derivation of boundary conditions 
(\ref{eq:ineq5}) was based on the assumption
that the damage zone expands and thus is always surrounded by an undamaged material. 
Numerically constructed damage profiles obtained by solving \eqref{dmg_non_lin}
with conditions (\ref{eq:ineq5}) are plotted in \figref{bif_shp}(b). 
The solution of the
first inelastic step is admissible and agrees well with the analytical
result~(\ref{eq:onset_solution}). In subsequent steps the size of the damage zone $L_d$
tends to decrease, but then the numerically computed damage would, at some points
near the boundary of the damage zone, decrease as well,
which  violates the damage irreversibility constraint. Such solutions are inadmissible
and
conditions (\ref{eq:ineq5})
need to be revisited. 

If $L_{d,k}<L_{d,k-1}$, then the damage at the boundary of $\Omega_{d,k}$
is not zero---it should be equal to the value after the previous step. 
This can be described by conditions
\beq\label{cond16}
\dmg_k(-L_{d,k}/2) = \dmg_{k-1}(-L_{d,k}/2), \hskip 10mm \dmg_k(L_{d,k}/2) = \dmg_{k-1}(L_{d,k}/2)
\eeq
For the spatial derivative of damage at $x=L_{d,k}/2$, 
we can again use condition (\ref{eq:ineq}) derived from the 
variational inequality (\ref{eq:optim_cond}), taking into account that $\dmg_k(x)=\dmg_{k-1}(x)$ for all $x\ge L_{d,k}/2$.
Consequently, (\ref{eq:ineq}) can be rewritten as
\beq\label{cond17}
\lim_{x \rightarrow L_{d,k}^-/2} \dmg'_k(x)
\geq
\dmg'_{k-1}(L_{d,k}/2)
\eeq
However, since $\dmg_k(L_{d,k}/2) = \dmg_{k-1}(L_{d,k}/2)$ and $\dmg_k(x) \ge \dmg_{k-1}(x)$ for
all $x<L_{d,k}/2$, condition (\ref{cond17}) can be satisfied only as an equality, and the damage 
profile remains continuously differentiable. Analogous arguments can be used at the left boundary
of the damage zone. The resulting conditions for the derivative of damage are
\beq\label{cond18}
\dmg_k'(-L_{d,k}/2) = \dmg_{k-1}'(-L_{d,k}/2), \hskip 10mm \dmg_k'(L_{d,k}/2) = \dmg_{k-1}'(L_{d,k}/2)
\eeq 

\begin{figure}[ht]
\begin{tabular}{cc}
\includegraphics{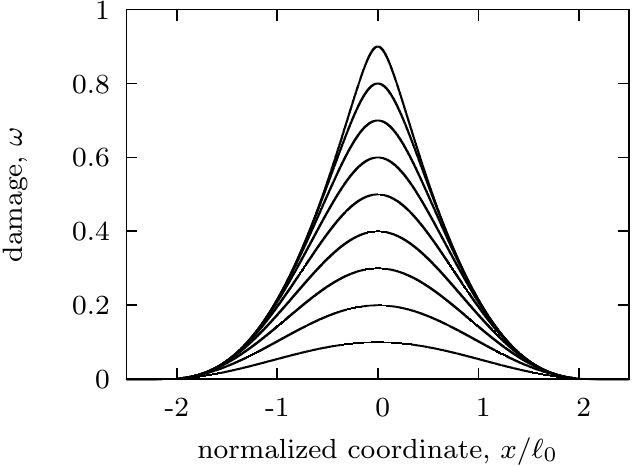}
&
\includegraphics{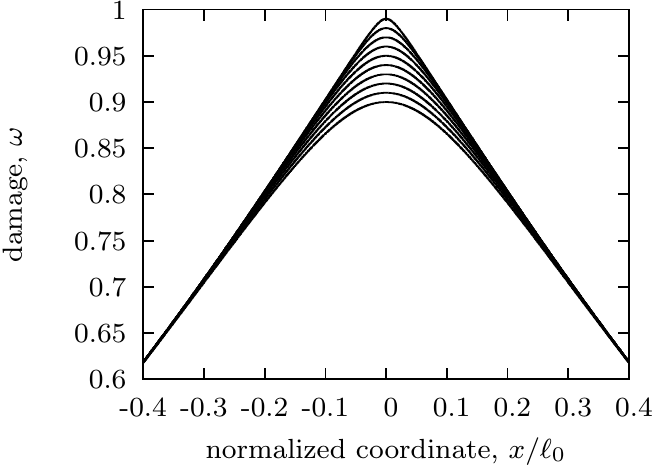}
\end{tabular}
\caption{Evolution of damage profile computed using conditions    (\ref{cond16}) and (\ref{cond18}) in
terms of the damage increment ($\Delta{\omega}=0$ and $\Delta{\omega}'=0$ at the boundary of the damage zone),
plotted (a) for maximum
damage $\omega_{\max}$ between 0.1 and 0.9, (b) for maximum damage  $\omega_{\max}$ between 0.9 and 0.99.}
\label{fig:uni_bar_damage}
\end{figure}

Imposing conditions (\ref{cond18}) combined with (\ref{cond16}),
we obtain an admissible damage
evolution shown in \figref{uni_bar_damage}, in which the size of the active part of the damage zone is
decreasing in time but the irreversibility constraint is satisfied and previously
generated damage does not decrease at any point. The solutions also meet the $C^1$ continuity inside the
damaged region, \eqref{EL3}, even for high values of damage.
The plotted profiles correspond to values of maximum damage $\omega_{\max}$ ranging from 0.1 to 0.9 with step 0.1,
but they have been constructed by an incremental numerical procedure with a much smaller step size,
making sure that the numerical error is negligible.


Interestingly, conditions (\ref{cond16}) and (\ref{cond18}) can be rewritten in terms of the damage
increments as
\bea
\Delta\dmg_k(-L_{d,k}/2) = \Delta\dmg_k(L_{d,k}/2) &=& 0\\
\Delta\dmg_k'(-L_{d,k}/2) = \Delta\dmg_k'(L_{d,k}/2) &=& 0
\eea
and they remain applicable even if the damage zone expands. However, if the problem is considered in the rate form,
it is not always correct to construct the appropriate conditions simply by replacing the damage increment by
damage rate. To gain more insight into the conditions that describe an evolving boundary between the (active) damage zone
and the zone of elastic unloading, let us consider an incremental step from $t_k$ to $t_{k+1}=t_k+\Delta t$
and let us express the computed damage zone size at the end of the step as $L_{d,k+1}(\Delta t)$, marking
explicitly that it depends
on the size of the step. In a similar spirit, the computed damage at the end of the step can be denoted as
$\dmg_{k+1}(x,\Delta t)$.  

In the case of an {\em expanding} damage zone, the solution 
satisfies conditions (\ref{eq:ineq5}), which can now be rewritten as 
\bea\label{cond53}
\dmg_{k+1}(L_{d,k+1}(\Delta t)/2,\Delta t) &=& 0 \\
\dmg_{k+1}'(L_{d,k+1}(\Delta t)/2,\Delta t) &=& 0 \label{cond54}
\eea
Since these conditions are satisfied for any step size $\Delta t$, we can
differentiate with respect to $\Delta t$ and evaluate the derivative at $\Delta t=0$, which yields
\bea\label{cond55}
\dmg_{k}'(L_{d,k}/2)\dot{L}_{d,k+1}(0)/2+\dot{\dmg}_k(L_{d,k}/2) &=& 0 \\
\dmg_{k}''(L_{d,k}/2)\dot{L}_{d,k+1}(0)/2+\dot{\dmg}_k'(L_{d,k}/2) &=& 0 \label{cond56}
\eea
Since $\dmg_{k}'(L_{d,k}/2)=0$, condition (\ref{cond55}) implies $\dot{\dmg}_k(L_{d,k}/2)=0$.
On the other hand, $\dmg_{k}''(L_{d,k}/2)$ is zero only for $k=1$; in general it is equal to 
$(1-\mu_k)/\ell_0^2$, as follows from \eqref{dmg_non_lin}. Therefore, it was correct to use 
conditions (\ref{bcondrate})
for the initial damage rate just at the onset of damage, but for $k>1$ we have $\mu_k<1$ and 
$\dmg_{k}''(L_{d,k}/2)=(1-\mu_k)/\ell_0^2>0$. Condition (\ref{cond56}) can then be used
to express the rate of the damage zone size as
\beq
\dot{L}_{d,k+1}(0) = -\frac{2\ell_0^2\dot{\dmg}_k'(L_{d,k}/2)}{1-\mu_k}
\eeq

As we can see, $\dot{\dmg}_k'(L_{d,k}/2)$ would not be zero if the damage zone truly expanded.
But for a {\em contracting} damage zone we have to use conditions (\ref{cond16}) and (\ref{cond18}),
and then (\ref{cond53})--(\ref{cond54}) is replaced by
\bea\label{cond57}
\dmg_{k+1}(L_{d,k+1}(\Delta t)/2,\Delta t) &=& \dmg_{k}(L_{d,k+1}(\Delta t)/2) \\
\dmg_{k+1}'(L_{d,k+1}(\Delta t)/2,\Delta t) &=& \dmg_{k}'(L_{d,k+1}(\Delta t)/2) \label{cond58}
\eea
and differentiation with respect to $\Delta t$ yields
\bea\label{cond59}
\dmg_{k}'(L_{d,k}/2)\dot{L}_{d,k+1}(0)/2+\dot{\dmg}_k(L_{d,k}/2) &=&\dmg_{k}'(L_{d,k}/2)\dot{L}_{d,k+1}(0)/2  \\
\dmg_{k}''(L_{d,k}/2)\dot{L}_{d,k+1}(0)/2+\dot{\dmg}_k'(L_{d,k}/2) &=& \dmg_{k}''(L_{d,k}/2)\dot{L}_{d,k+1}(0)/2 \label{cond60}
\eea
which implies
\beq
\dot{\dmg}_k(L_{d,k}/2)=0, \hskip 10mm \dot{\dmg}_k'(L_{d,k}/2)=0
\eeq
This means that, in the case of a contracting zone, conditions (\ref{bcondrate}) can be used for any $k$.

\begin{figure}[h]
\begin{tabular}{cc}
\includegraphics{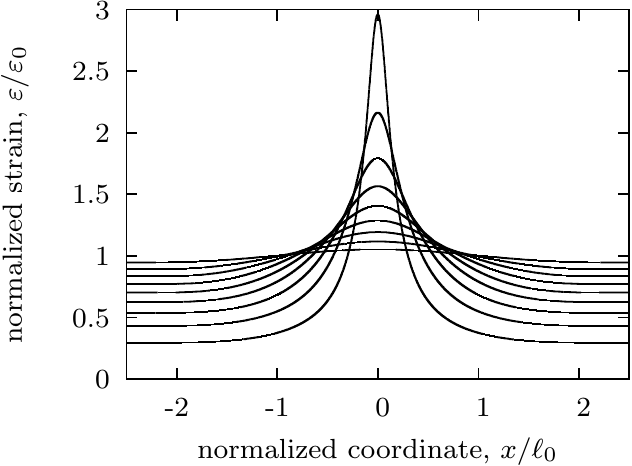}
&
\includegraphics{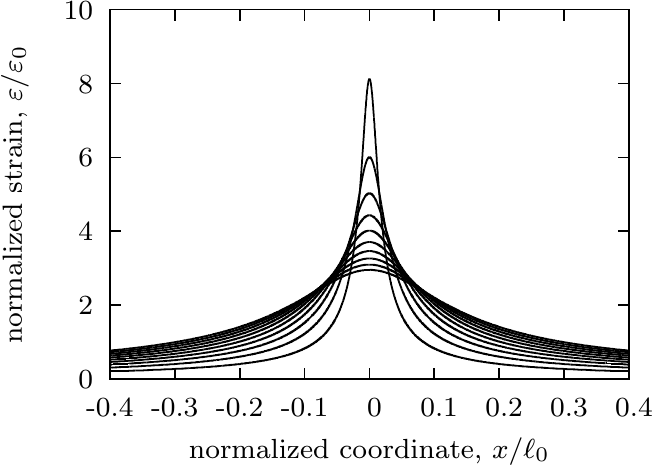}
\end{tabular}
\caption{Evolution of strain profile computed using conditions (\ref{cond16}) and (\ref{cond18}) in
terms of the damage increment ($\Delta{\omega}=0$ and $\Delta{\omega}'=0$), plotted (a) for maximum damage $\omega_{\max}$ between 0.1 and 0.9,
(b) for maximum damage $\omega_{\max}$ between 0.9 and 0.99.}
\label{fig:uni_bar_strain}
\end{figure}

After this detailed discussion of conditions imposed at the evolving boundary of the damage zone,
let us turn our attention back to the numerically computed solution of \eqref{dmg_non_lin}.
The damage profiles from \figref{uni_bar_damage} are complemented by the evolution of strain
profiles obtained from \eqref{sigma_const} and normalized by the limit elastic
strain,  $\eps_0=\sqrt{2\gf / E}$; see~\figref{uni_bar_strain}. The results
confirm that not only damage but also strain tends to localize into a contracting zone, while
the strains in the elastically unloading zones are decreasing.

\begin{figure}[h]
\begin{tabular}{cc}
\includegraphics{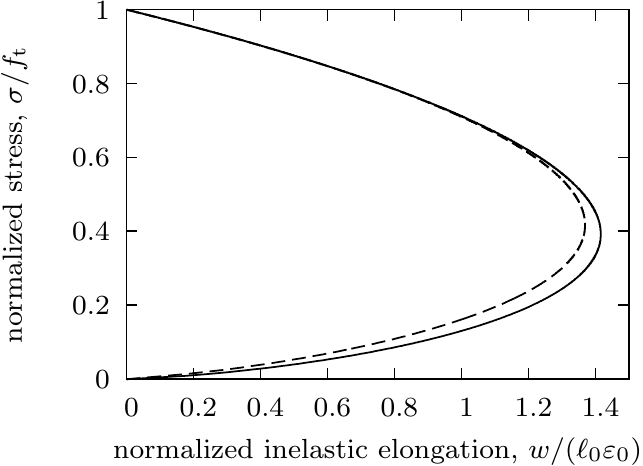}
&
\includegraphics{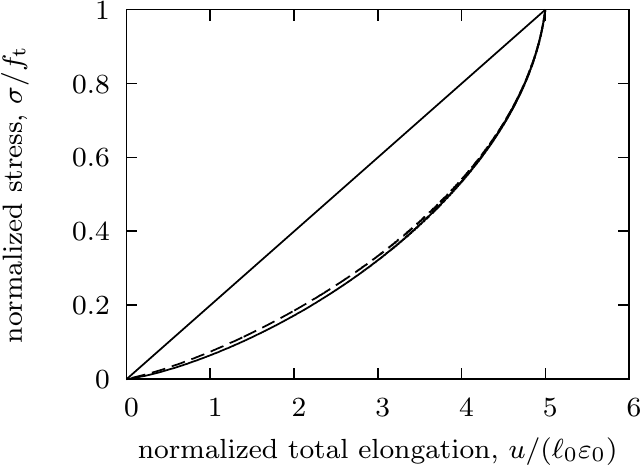}
\end{tabular}
\caption{Relation between stress and (a) inelastic part of elongation, (b) total elongation (for a bar of total length $L=5\ell_0$); solid
curves correspond to conditions (\ref{cond16}) and (\ref{cond18}) in
terms of the damage increment ($\Delta{\omega}=0$ and $\Delta{\omega}'=0$), 
dashed curves to conditions  (\ref{eq:ineq5}) in terms of the total
damage ($\dmg=0$ and $\dmg'=0$).} 
\label{fig:uni_bar_diagram_structure}
\end{figure}

In order to assess the implication of the strain localization at the
structural level, we also characterize the response by the inelastic
part of the elongation, obtained as
\bea\nonumber
w_{k} 
& = & 
\int_{-L/2}^{L/2} 
\left(\eps_k(x) - \frac{\sigma_k}{E}\right)\,\dx 
=
\int_{-L/2}^{L/2} 
\left(
  \frac{\sigma_k}{E(1-\omega_k(x))} 
  - 
  \frac{\sigma_k}{E}
\right)\dx =
\\
&=&
\frac{\sigma_k}{E}
\int_{-L/2}^{L/2} 
 \frac{\omega_k(x)}{1-\omega_k(x)} \dx 
=
\frac{\sigma_k}{E}
\int_{-L_{d,\max}/2}^{L_{d,\max}/2} 
 \frac{\omega_k(x)}{1-\omega_k(x)} \dx 
\label{www}
\eea
where $L_{d,\max}$ is the maximum size of the damage zone ever reached in the
previous history (in the present case of a contracting zone it is equal to $L_{d,2}$). 
Again, a dimensionless representation is used, in which the inelastic
displacement is normalized by $\ell_0\eps_0$ and the stress
$\sigma_k$ is divided by the tensile strength, $\ft=\sqrt{2E\gf}$. The
normalized  diagram in \figref{uni_bar_diagram_structure}(a) reveals that the
 inelastic displacement
reaches its maximum after softening to $\approx 40\%$ of the peak stress and then 
decreases to zero. In terms of the structural response, represented by
the diagram of stress versus total elongation, this leads to a strong snapback;
see \figref{uni_bar_diagram_structure}(b). The total elongation is easily expressed
as
\beq
u_{k} 
 = 
\int_{-L/2}^{L/2} \eps_k(x)\,\dx = w_k + \frac{\sigma_k L}{E}
\eeq
and in \figref{uni_bar_diagram_structure}(b) it is normalized by $\ell_0\eps_0$.
Of course, snapback at the structural level can always be expected for very long 
bars, but here it occurs for short bars, too. The diagram has been plotted for the total bar length set to $L=5\ell_0$, which is only slightly larger than 
the initial damage zone size $L_{d,2}=\sqrt{2}\pi\ell_0\approx 4.443\ell_0$. 

\begin{figure}[ht]
\begin{center}
\includegraphics{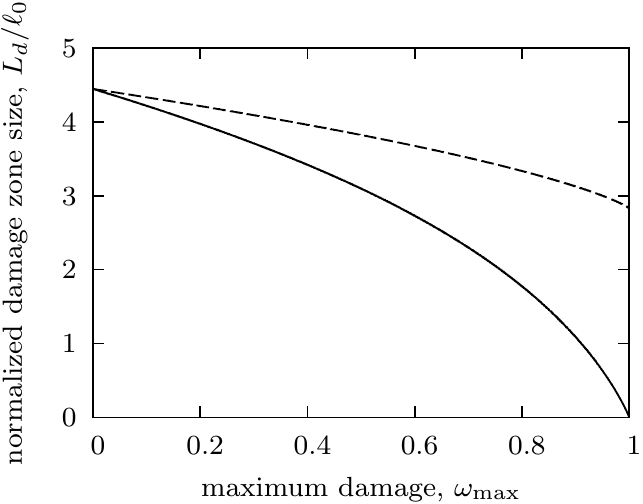}
\end{center}
\caption{Evolution of the size of damage zone as a function of maximum damage; solid
curves correspond to conditions (\ref{cond16}) and (\ref{cond18}) in
terms of the damage increment ($\Delta{\omega}=0$ and $\Delta{\omega}'=0$), 
dashed curves to conditions  (\ref{eq:ineq5}) in terms of the total
damage ($\dmg=0$ and $\dmg'=0$).} 
\label{fig:uni_bar_diagram_structure2}
\end{figure}

Shrinking of the damage zone is further documented in 
\figref{uni_bar_diagram_structure2}, which shows the evolution of $L_{d}$ and demonstrates that,
at complete failure, the active zone
degenerates to a point. It is interesting to observe
that when the (incorrect) boundary conditions in terms of the total damage are
applied, the damaging zone is substantially larger and is of a non-zero length
at failure; see the dashed curve in \figref{uni_bar_diagram_structure}(b).
Still, this phenomenon has limited effect on the structural
response, and is not strong enough to eliminate the snapback phenomenon; see the
dashed curves in \figref{uni_bar_diagram_structure}.

In the following sections, we shall investigate to which extent such behavior
can be influenced by more advanced constitutive models. To avoid a
profusion of notation, the indices referring to a given time step are omitted in
what follows.

\section{Generalized formulation with softening}\label{sec4}

\subsection{Governing equations}\label{sec4.1}

The model considered in the previous section exhibits an extremely brittle behavior.
This is related to the specific definition of the dissipation distance (\ref{eq:GDD}),
which is motivated by the concept of an elastic-brittle response. In general it can be replaced by 
\begin{equation}\label{eq:GDDm}
\DD( \tdmg_1, \tdmg_2 )
=
\left\{ 
\begin{array}{ll}
\displaystyle \int_\dmn \left( \Diss(\tdmg_2(x)) - \Diss(\tdmg_1(x)) \right) \de x
  & \mbox{if } \tdmg_2 \geq \tdmg_1 \mbox{ in }
  \dmn \\ +\infty & \mbox{otherwise}
\end{array}
\right.
\end{equation}
where $\Diss$ is a function that represents the density of energy dissipated
in a damage process up to a given damage level. This function should be non-decreasing,
with $\Diss(0)=0$ and $\Diss(1)=\gff=$ density of energy dissipated by the complete failure process. 

\begin{figure}[ht]
\centering
\includegraphics{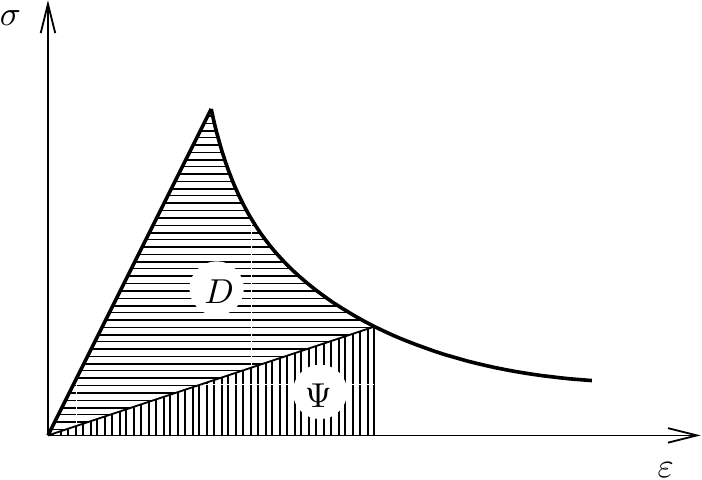} 
\caption{Stress-strain diagram (under uniform strain) and the meaning of free energy $\Psi$ and dissipated energy $\Diss$.}\label{fig:f_diagram}
\end{figure}

To construct simple but physically meaningful forms of function $\Diss$, let us consider
the idealized case of uniform damage, such that the regularizing part of stored
energy \rev{remains} equal to zero. The stored energy then reduces to
the standard part (\ref{eq:GEstd}), which can be rewritten as
\begin{equation}
\EF_{\rm std}(\tdisp, \tdmg ) = \int_\dmn \EE(\tdisp'^2(x),\tdmg(x))  \de x
\end{equation}
where
\beq
\EE(\eps,\omega) = \half(1-\omega)E\eps^2
\eeq
is the density of stored elastic energy (more precisely, of the Helmholtz
free energy). The graphical meaning is illustrated
in \figref{f_diagram}. The vertically hashed area corresponds to the density of free energy
and the horizontally hashed area is the density of dissipated energy, $\Diss(\omega)$. Under isothermal
conditions, their
sum must be equal to the density of supplied work (all densities being taken per unit volume). 

From the incremental energy balance equation
\beq\label{eq48}
\frac{\partial\EE}{\partial\eps}{\rm d}\eps + \frac{\partial\EE}{\partial\omega}{\rm d}\omega +
\frac{{\rm d}\Diss}{{\rm d}\omega}{\rm d}\omega = \sigma\,{\rm d}\eps
\eeq
we get the stress-strain relation
\beq
\sigma = \frac{\partial\EE}{\partial\eps} = (1-\omega)E\eps
\eeq
and an additional equation
\beq\label{eq:mj1}
\frac{{\rm d}\Diss}{{\rm d}\omega} = - \frac{\partial\EE}{\partial\omega} = \half E\eps^2
\eeq
in which $\eps$ has to be understood as the strain that produced the current damage level $\omega$.
The negative derivative of free energy density with respect to the damage variable is 
called the damage energy release rate and denoted as $Y$.

Suppose that we would like to construct a model which, under uniform damage and monotonic loading, 
produces a specific shape of the stress-strain diagram, described by
\beq
\sigma = s(\eps)
\eeq
where $s$ is a given function. To recover this stress-strain relation from the 
constitutive law of the damage model,
\beq
\sigma = (1-\omega)E\eps
\eeq
we need to consider the damage variable as a function of the current strain (still assuming monotonic loading) and
postulate a damage law in the form
\beq\label{eq:dam1}
\omega = 1 - \frac{s(\eps)}{E\eps}\equiv
g(\eps)
\eeq
In this way, the damage function $g$ can be constructed directly from the given shape of the
stress-strain diagram. In the interval of growing damage (i.e., above the damage threshold
and below the state of complete damage), the damage law is invertible and the strain $\eps$ corresponding
to a given damage level $\omega$ can be expressed as
\beq\label{eq:dam1x}
\eps = g^*(\omega)
\eeq
where $g^*$ is the inverse function of $g$. Recall that the damage energy release rate, $-\partial\Psi/\partial\omega$,
has been denoted as $Y$. 
Consequently, \eqref{mj1} can be rewritten as
\beq\label{eq:mj2}
\frac{\mbox{d}\Diss(\omega)}{\mbox{d}\omega}=Y(\omega) =\half Eg^{*2}(\omega)
\eeq
By integrating this relation, we can construct an appropriate formula for the dissipation
density function
\beq\label{eq:mj2b}
\Diss(\omega) = \int_0^{\omega} Y(\widetilde\omega)\de \widetilde\omega=
\half E\int_0^{\omega} g^{*2}(\widetilde\omega)\de \widetilde\omega
\eeq

Let us emphasize that Eqs.~(\ref{eq48})--(\ref{eq:mj2b}) apply to the local version
of the damage model, with regularizing terms omitted. For such a local model, the local values
of strain and damage at a given point are, during damage growth, directly linked by \eqref{dam1}. 
However, for a model regularized by the energy term dependent on the damage gradient,
the analysis from Section~\ref{sec:sec3} must be repeated with an appropriate modification---the 
term $\delta\omega(x)\gf$ in \eqref{mj3} is replaced
by $\delta\omega(x)\DissD(\omega(x))$ while the regularizing term remains unchanged.
The modified form of \eqref{mj4} is then 
\begin{eqnarray}\nonumber
-
\int_\dmn
\vdisp(x) 
\Bigl( ( 1 - \dmg(x) ) E \disp'(x) \Bigr)'
\de x
-
\\ \nonumber
-
\int_\dmn
\vdmg(x)
\left(
\half
 E \disp'^2(x)
+
\gfn \il \dmg''(x)
-
\DissD(\omega(x))
\right)
\de x 
+
\\
+
\left[ \vdmg(x) \gfn \il \dmg'(x) \right]_{\bnd}
& \geq & 0
\label{eq:mj5}
\end{eqnarray}
Inside the damage zone, the variation $\vdmg(x)$ is arbitrary and the resulting optimality condition
\beq\label{eq:mj5x}
\half
 E \disp'^2(x)
+
\gfn \il \dmg''(x)
-
\DissD(\omega(x))=0
\eeq
has the form of a differential equation. Repeating the procedure from Section~\ref{sec:elastic-brittle-governing},
we can convert it into 
\begin{equation}\label{eq:mj7}
\il \dmg''(x) + \frac{\mu}{( 1 - \dmg(x) )^2} = \frac{\DissD(\omega(x))}{\gfn}
\end{equation}
which is the generalized form of \eqref{dmg_non_lin}, to be used in subsequent analyses.
However, it is also instructive to look at another interpretation of \eqref{mj5x}.
Taking into account \eqref{mj2} and denoting $\disp'$ as $\eps$, we can rewrite \eqref{mj5x} as
\beq\label{eq:mj91}
\eps^2(x) = g^{*2}(\dmg(x))-\frac{2\gfn}{E}\il\dmg''(x)
\eeq
This is a differential relation between strain and damage that is valid inside the damage zone and replaces
the algebraic relation (\ref{eq:dam1x}). Since $\sqrt{2\gfn/E}=\eps_0=$ limit elastic strain, we can further rewrite
\eqref{mj91} as
\beq
\eps(x) = \sqrt{g^{*2}(\dmg(x))-\eps_0^2\il\dmg''(x)}
\eeq
or, in the inverse form, as
\beq
\dmg(x) = g\left(\sqrt{\eps^2(x)+\eps_0^2\il\dmg''(x)}\right)
\eeq
The last relation is a generalized form of \eqref{dam1}. It shows that the regularizing term accelerates 
the damage growth (as compared to the local model) in regions where the current damage profile is convex
($\dmg''(x)>0$)  and decelerates it in regions where the current damage profile is concave ($\dmg''(x)<0$).
This, combined with continuity of damage, prevents the damage profile from becoming ``too much localized''.

\subsection{Model with linear softening}

The simplest description of a stress-strain law with softening is based on the
linear relation between stress and strain in the post-peak range, from the
limit elastic strain $\eps_0$ to a certain failure strain $\eps_f$, which must not
be smaller than $\eps_0$. The special case of $\eps_f=\eps_0$ corresponds to the
brittle model considered in the previous section, and for increasing $\eps_f$ we obtain
a more ductile behavior. Therefore, the dimensionless ratio 
\beq\label{eq:beta}
\beta=\frac{\eps_0}{\eps_f}
\eeq
 which is between 0 and 1,
can be called the brittleness number.

The softening branch of the stress-strain diagram with linear softening is described by
\beq
\sigma = s(\eps) \equiv E\eps_0 \frac{\eps_f-\eps}{\eps_f-\eps_0}, \hskip 10mm \eps_0\le \eps \le\eps_f
\eeq
and the corresponding damage function evaluated according to \eqref{dam1} is
\beq
g(\eps) = 1 - \frac{s(\eps)}{E\eps} =    1 - \frac{\eps_0}{\eps}\cdot\frac{\eps_f-\eps}{\eps_f-\eps_0}=
\frac{(\eps-\eps_0)\eps_f}{(\eps_f-\eps_0)\eps}=\frac{1}{1-\beta}\left(1-\frac{\eps_0}{\eps}\right), \hskip 10mm \eps_0\le \eps \le\eps_f
\eeq
In this simple case, the inverse function can be evaluated analytically as
\beq\label{eq:mj8}
g^*(\omega) =\frac{\eps_0}{1-\omega+\beta\omega}, \hskip 10mm 0\le \omega\le 1
\eeq
and after substitution into (\ref{eq:mj2})--(\ref{eq:mj2b})  we get
\begin{eqnarray}\label{eq:dissd1}
\DissD(\omega) &=& \half Eg^{*2}(\omega) = \half E\frac{\eps_0^2}{(1-\omega+\beta\omega)^2}= \frac{\gfn}{(1-\omega+\beta\omega)^2}
\\
\Diss(\omega) &=&\int_0^{\omega} Y(\widetilde\omega)\de \widetilde\omega = 
\frac{\gfn}{1-\beta}\left(\frac{1}{1-\omega+\beta\omega}-1\right)
\end{eqnarray}
For $\omega=1$, the last expression gives 
\beq\label{eq:mj9}
\gff = \Diss(1) = 
\frac{\gfn}{1-\beta}\left(\frac{1}{\beta}-1\right) = \frac{\gfn}{\beta}
\eeq
Recall that $\beta=\eps_0/\eps_f=$ brittleness number between 0 and 1.
The case of $\beta=1$ corresponds
to the elastic-brittle model from Section~\ref{sec4}, smaller values
correspond to a less brittle behavior, with $\gff > \gfn$.
\eqref{mj7}, to be satisfied along the damage zone, is for the
case of linear softening written as  
\begin{equation}\label{eq:mj7b}
\il \dmg''(x) + \frac{\mu}{( 1 - \dmg(x) )^2} = \frac{1}{(1-\omega(x)+\beta\omega(x))^2}
\end{equation}

\begin{figure}[p]
\begin{tabular}{cc}
(a) & (b)
\\
\includegraphics{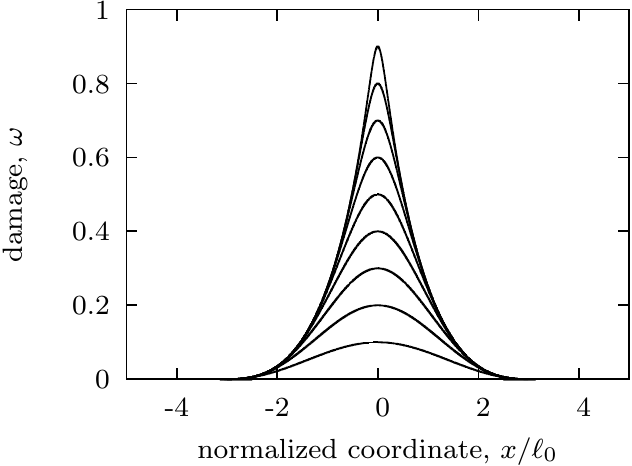}
&
\includegraphics{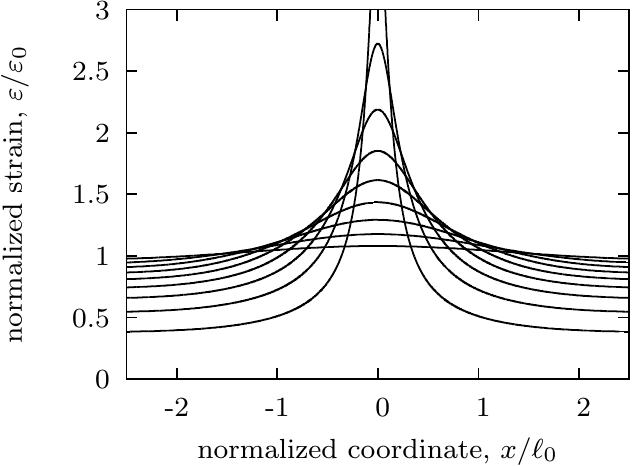}
\end{tabular}
\caption{Linear softening: profiles of (a)  damage and (b) strain, for $\beta=0.5$
and $\omega_{\max}$ between 0.1 and 0.9.}
\label{fig:ls1}
\begin{tabular}{cc}
(a) & (b)
\\
\includegraphics{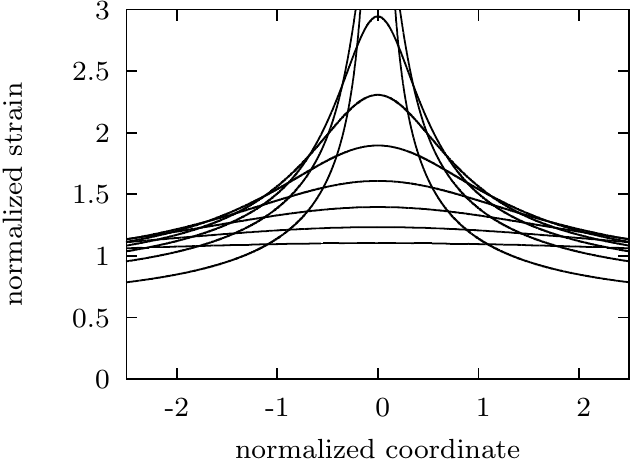}
&
\includegraphics{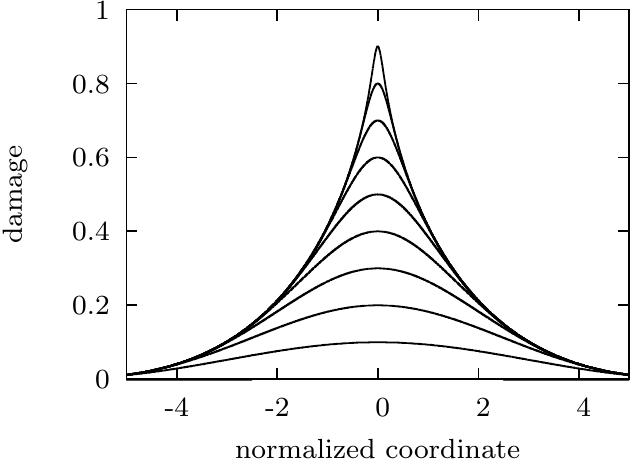}
\end{tabular}
\caption{Linear softening: profiles of (a)  damage and (b) strain, for $\beta=0.1$
and $\omega_{\max}$ between 0.1 and 0.9.}
\label{fig:ls2}
\centering
\includegraphics{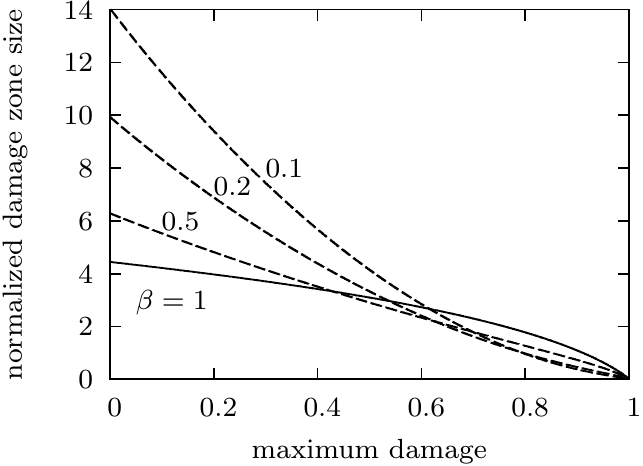}
\caption{Linear softening: 
size of the active part of damage zone plotted as a function of maximum damage for different values of brittleness number $\beta$.}
\label{fig:ls3}
\end{figure}

\begin{figure}[h]
\begin{tabular}{cc}
(a) & (b)
\\
\includegraphics{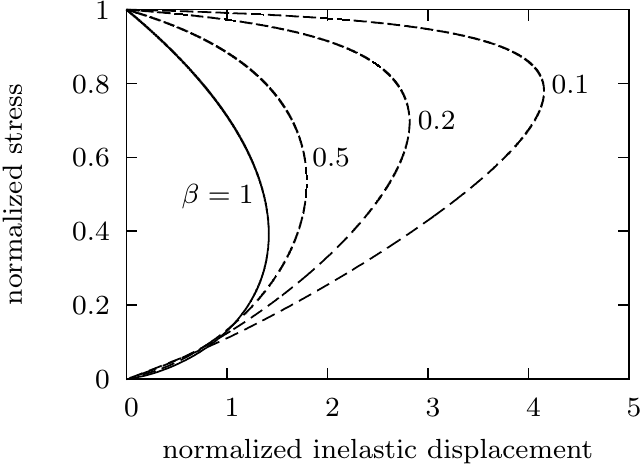}
&
\includegraphics{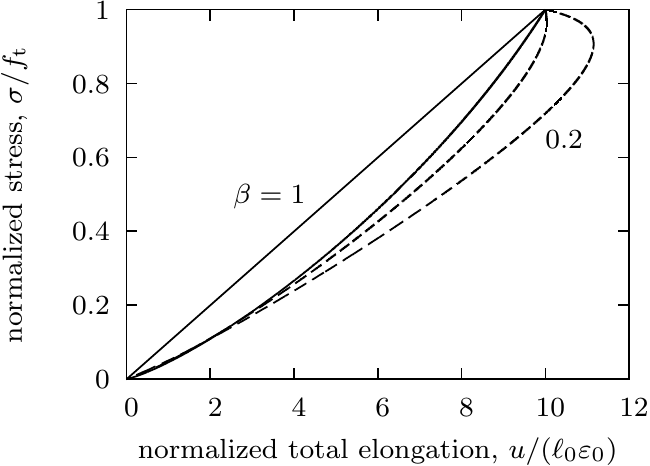}
\end{tabular}
\caption{Linear softening: (a) relation between stress and inelastic part of elongation, (b) relation between stress and total elongation (plotted for bar length $L=10\ell_0$), both for different values of brittleness number $\beta$.}
\label{fig:ls4}
\end{figure}

The numerically evaluated evolution of the damage and strain profiles is shown in \figref{ls1} for $\beta=0.5$
and in \figref{ls2} for $\beta=0.1$. 
For a material with more ductile local response
(lower brittleness number, \figref{ls2}), the initial size of the damage zone is larger, but the 
active part of the damage zone is shrinking and tends to one single cross section of the bar
as the maximum damage tends to 1. This is shown in \figref{ls3}, in which the solid curve corresponds to 
the elastic-brittle model ($\beta=1$) and dashed curves to the model with linear softening and 
brittleness numbers $\beta = 0.5$, 0.2 and 0.1. The resulting load-displacement diagrams are shown
in \figref{ls4}, first with the inelastic part of elongation on the horizontal axis (\figref{ls4}a),
and then with the total elongation on the horizontal axis (\figref{ls4}b).
Here, the bar length was set to $L=10\ell_0$, which is needed to make sure that the localized solution obtained
for $\beta=0.2$ is admissible. For $\beta=0.1$ (not covered in \figref{ls4}b), a still longer bar would be
needed, because the initial size of the damage zone is about $14\ell_0$; see the top curve in \figref{ls3}.

\subsection{Model with exponential softening}

For concrete and similar quasi-brittle materials, damage laws that lead to a long tail of the softening
curve are often used. In the local setting, a suitable shape of the stress-strain diagram is obtained 
e.g.\ with the exponential softening law,
\beq
\sigma = s(\eps) \equiv E\eps_0\exp\left(-\frac{\eps-\eps_0}{\eps_f}\right), \hskip 10mm \eps_0\le\eps
\eeq
from which
\beq
g(\eps) = 1-\frac{s(\eps)}{E\eps} = 1- \frac{\eps_0}{\eps} \exp\left(-\frac{\eps-\eps_0}{\eps_f}\right), \hskip 10mm \eps_0\le\eps
\eeq
In this case, the relation $\omega=g(\eps)$ cannot be inverted in closed form, but the inverse function $g^*$
can be defined implicitly by the equation
\beq\label{eq:inver}
(1-\omega) g^*(\omega)\exp\left(\frac{g^*(\omega)-\eps_0}{\eps_f}\right) = \eps_0, \hskip 10mm 0\le \omega\le 1
\eeq
The corresponding values of $\DissD(\omega)$ can then be evaluated by simple substitution into \eqref{mj2}.
Note that function $\Diss$, which would have to be constructed by numerical integration,
is not really needed. This function was useful for the formal derivation but it does not
appear in the resulting  \eqref{mj7} directly, only through its first derivative $\DissD$.

\begin{figure}[p]
\begin{tabular}{cc}
(a) & (b)
\\
\includegraphics{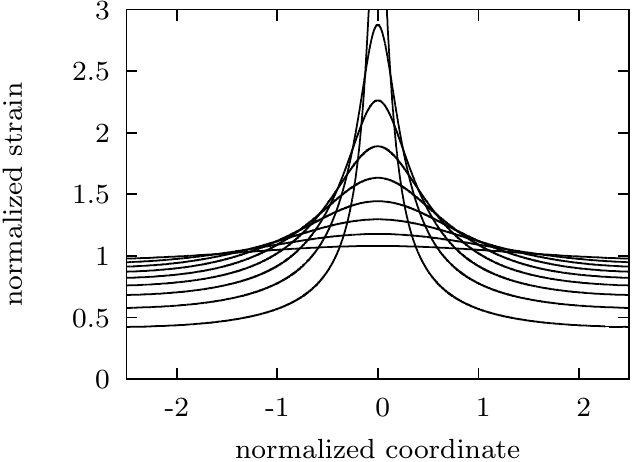}
&
\includegraphics{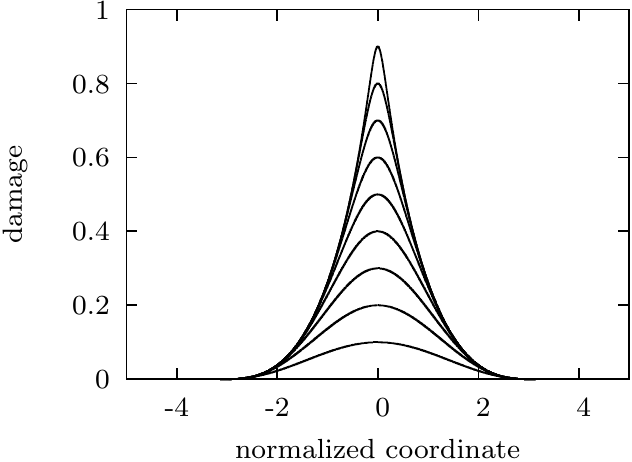}
\end{tabular}
\caption{Exponential softening: profiles of (a) strain, (b) damage; all for $\beta=0.5$.}
\label{fig:es1}
\begin{tabular}{cc}
(a) & (b)
\\
\includegraphics{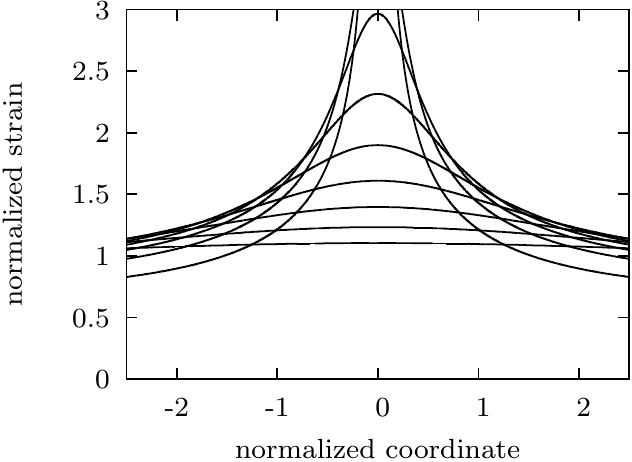}
&
\includegraphics{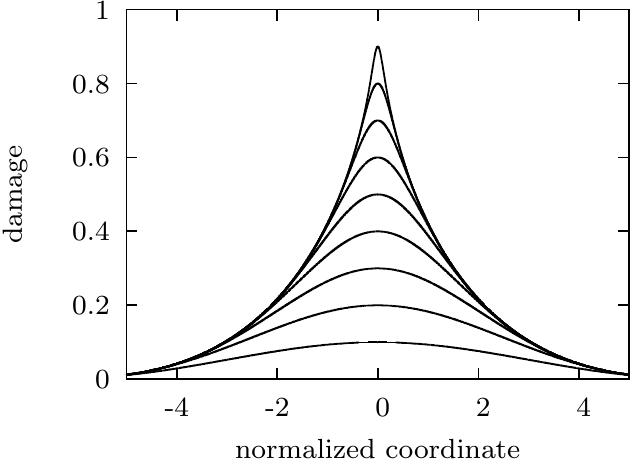}
\end{tabular}
\caption{Exponential softening: profiles of (a) strain, (b) damage; all for $\beta=0.1$.}
\label{fig:es2}
\bigskip
\centering
\includegraphics{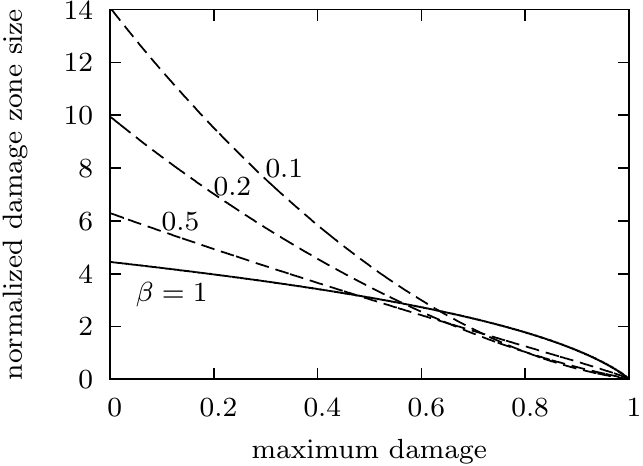}
\caption{Exponential softening: evolution of the size of damage zone (active part) as a function of maximum damage; 
solid curve corresponds to elastic-brittle model, dashed curves to exponential softening with $\beta=0.5$, 0.2 and 0.1.}
\label{fig:es3}
\end{figure}

Evolution of the strain and damage profiles is shown in \figref{es1} for $\beta=0.5$
and in \figref{es2} for $\beta=0.1$.
Even for exponential softening and a low brittleness number, the active part of the damage zone is shrinking.
As seen in \figref{es3}, the active part of the damage zone approaches a single cross section as the maximum damage
tends to 1.  This happens for all values of the brittleness number. A low brittleness number $\beta$ leads to
a large initial size of the damage zone but at later stages of the localization process the influence of $\beta$
fades away.
 
\begin{figure}[h]
\begin{tabular}{cc}
(a) & (b)
\\
\includegraphics{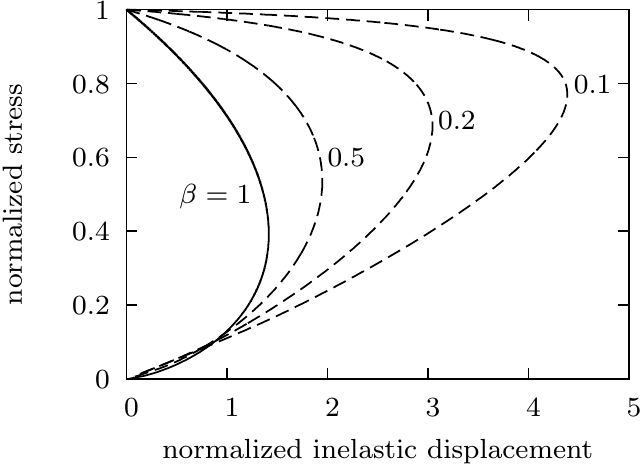}
&
\includegraphics{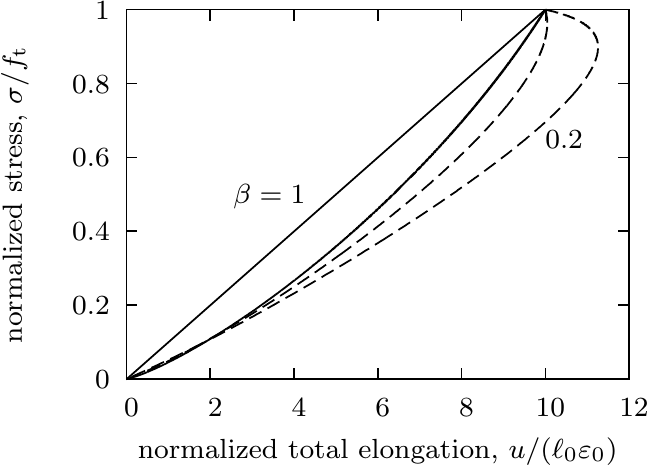}
\end{tabular}
\caption{Exponential softening:  (a) relation between stress and inelastic part of elongation, (b) relation between stress and total elongation (plotted for bar length $L=10\ell_0$); solid curve corresponds to elastic-brittle model, dashed curves to exponential softening with different values of brittleness number $\beta$.}
\label{fig:es4}
\end{figure}

The load-displacement diagrams in \figref{es4} have a similar shape to those corresponding to a model with linear softening.
Parameter $\beta$ affects the initial post-peak slope of the diagram and the amount of dissipated energy,
but even for low values of $\beta$ the diagrams exhibit snapback and eventually return to the origin.
Therefore, it seems that an adjustment of the dependence of dissipated energy density on damage is not sufficient 
to construct a model that exhibits a long tail of the load-displacement diagram. Another modification of the model
equations is needed. 

\section{Modified regularization techniques for elastic-brittle models}\label{sec5}

\subsection{Approach based on gradient of inelastic compliance variable}

The results obtained in the previous section indicate that even if the damage law is constructed from a local model
with a long tail of the stress-strain diagram, such as the exponential softening model, the global load-displacement
diagram exhibits snapback and the behavior at late stages of the damage process is extremely brittle, which is acceptable
only for a certain limited class of materials. Typical load-displacement diagrams of 
quasi-brittle materials such as concrete possess a long tail. The absence of such a tail in our numerical results
is related to the dramatic shrinking of the active part of the damage zone. The primary reason is that the
regularizing term based on the damage gradient becomes less powerful at late stages of the process, because
the damage variable is bounded by its maximum value 1 and this limit value can be approached at the center of 
a shrinking damage zone without creating extremely steep damage gradients; see
e.g.~\figref{es2}b.

A remedy can be sought in a modification
of the regularizing term, based on a transformed variable that approaches infinity as damage approaches 1. 
A good candidate is the inelastic compliance variable, defined as
\beq\label{eq:mj21}
\gamma = \frac{\omega}{1-\omega}
\eeq
Its physical meaning can be explained based on the decomposition of strain into the elastic and inelastic parts:
\beq
\eps = \frac{\sigma}{(1-\omega)E} =   \frac{\sigma}{E}  +  \left(\frac{1}{1-\omega}-1\right)\frac{\sigma}{E}
=\frac{\sigma}{E}  +\frac{\omega}{1-\omega}\frac{\sigma}{E}
\eeq
The first term on the right-hand side is the elastic strain, $\sigma/E$, and the second term may be considered
as the inelastic
part of strain, equal to the elastic strain multiplied by the inelastic compliance variable $\gamma$ from \eqref{mj21}.
During the elastic stage of the response, $\gamma$ vanishes. As damage evolves and $\omega$ approaches 1, 
$\gamma$ grows to infinity, and its gradient can become very steep. Therefore, it can be expected that regularization
based on the gradient of $\gamma$ remains efficient even at late stages of the damage process.

Consider a modified model with  $\tdmg'(x)$ in \eqref{GEreg} replaced by $\widehat{\gamma}'(x)$. 
The regularizing term is now
\beq\label{eq:mj22}
{\cal E}_{\rm reg}(\tdmg) = \half\int_{\Omega} \gfn\il \left(\frac{\tdmg(x)}{1-\tdmg(x)}\right)'^2\,{\rm d}x=
\half\int_{\Omega} \gfn\il \frac{\tdmg'^2(x)}{(1-\tdmg(x))^4}\,{\rm d}x
\eeq
The dissipation distance is considered in its generalized form (\ref{eq:GDDm}), which corresponds to a model with softening. 
After appropriate modifications, the procedure from Sections \ref{sec:elastic-brittle-regularity} 
and \ref{sec4.1} leads to the governing equation
\beq
\il\omega''(x) +\frac{2\il}{1-\omega(x)}\omega'^2(x) +\mu(1-\omega(x))^2 = 
\frac{(1-\omega(x))^4}{\gfn}\DissD(\omega(x))
\eeq
which represents the modified form of \eqref{mj7}.

\begin{figure}[h]
\begin{tabular}{cc}
(a) & (b)
\\
\includegraphics{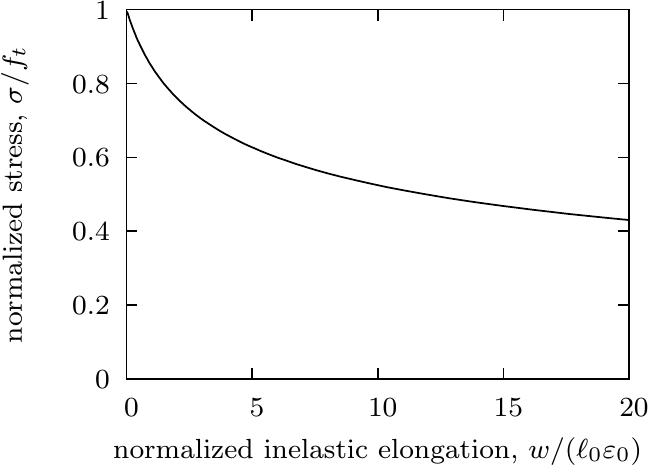}
&
\includegraphics{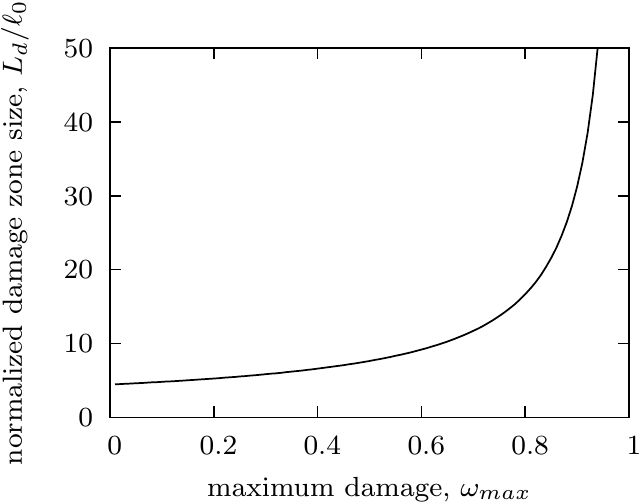}
\end{tabular}
\caption{Elastic-brittle model with gradient of inelastic compliance variable:  (a) relation between stress and inelastic part of elongation, (b) size of damage zone (active part) as a function of maximum damage.}
\label{fig:m1}
\end{figure}

\begin{figure}[h]
\begin{tabular}{cc}
(a) & (b)
\\
\includegraphics{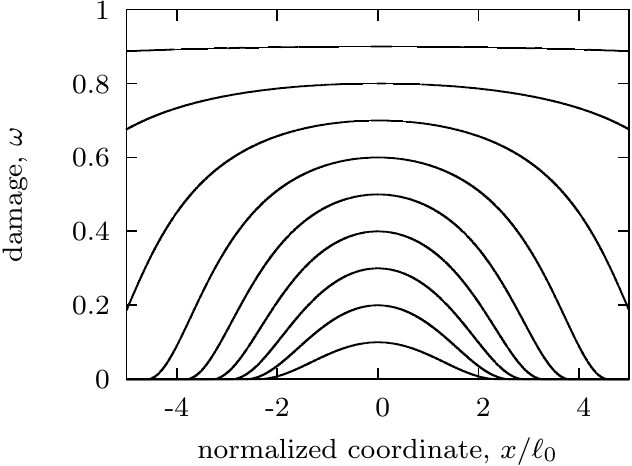}
&
\includegraphics{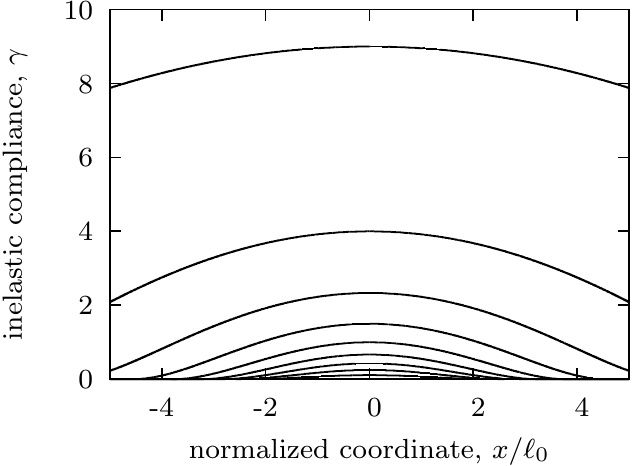}
\end{tabular}
\caption{Elastic-brittle model with gradient of inelastic compliance variable:  evolution of (a) damage profile,
(b) profile of inelastic compliance variable $\gamma$.}
\label{fig:m2}
\end{figure}

The numerical solution indeed leads to a long tail of the load-displacement diagram. It is not even necessary to
use a model with gradual softening, since the global response is quite ductile already for the locally
elastic-brittle model; see \figref{m1}a. 
Unfortunately, the ductile response is associated here with an expansion of the damage zone; see \figref{m1}b and \figref{m2}a.
This is clearly a pathological phenomenon, and the result for a model with linear or exponential softening would
be even worse. At late stages, the regularization becomes too strong. If damage is close to 1, small gradients of damage
correspond to large gradients of the inelastic compliance variable $\gamma$. This is clear from \figref{m2}b, which
shows the evolution of the profile of $\gamma$.

\subsection{Approach based on variable characteristic length}\label{sec5.2}

Further modification of the regularizing term can be inspired by an interpretation of 
\eqref{mj22} as
\beq\label{eq:mj23}
{\cal E}_{\rm reg}(\tdmg) = 
\half\int_{\Omega} \gfn\,\ell^2(\tdmg(x))\, \tdmg'^2(x)\,{\rm d}x
\eeq
The length parameter $\ell$ is now considered as variable, dependent on the current damage.
For the specific choice
\beq\label{eq:mj24}
\ell(\dmg)=\frac{\ell_0}{(1-\dmg)^2}
\eeq
we recover the model from Section~\ref{sec5.2}, but it is also possible to use other expressions.
The governing equation can be written in the general form 
\beq\label{eq:mj27}
\ell^2(\dmg(x))\,\omega''(x) +\ell(\dmg(x))\,\ello(\dmg(x))\,\omega'^2(x) +\frac{\mu}{(1-\omega(x))^2} = 
\frac{\DissD(\omega(x))}{\gfn}
\eeq
where $\ello\equiv{\rm d}\ell/{\rm d}\omega$.
To better control the regularizing effect, we can replace \eqref{mj24} for instance by
\beq\label{eq:mj25}
\ell(\dmg)=\frac{\ell_0}{(1-\dmg)^p}
\eeq
which leads to
\beq\label{eq:mj26}
\ello(\dmg)=\frac{{\rm d}\ell(\omega)}{{\rm d}\omega}=\frac{p\ell_0}{(1-\dmg)^{p+1}}
\eeq
The exponent $p$ is an adjustable parameter, with $p=0$ corresponding to the initial formulation
based on the gradient of damage (see Section~\ref{sec4})
and $p=2$  to the modified 
formulation based on the gradient of compliance variable (see Section~\ref{sec5.2}). 

Let us explore the influence of parameter $p$ on the response of the regularized elastic-brittle model, for which the right-hand
side of \eqref{mj27} is equal to~1. Substituting \eqref{mj25} and \eqref{mj26} into \eqref{mj27} with a unit right-hand side
and multiplying both sides by $(1-\omega)^{2p+1}$, we obtain the nonlinear second-order differential equation
\beq\label{eq:mj29}
(1-\omega(x))\ell_0^2\omega''(x) +p\ell_0^2\omega'^2(x) + (1-\omega(x))^{2p-1}\mu = (1-\omega(x))^{2p+1} 
\eeq
At the onset of damage, we have $\omega_1(x)=0$ and $\mu=1$, and the initial damage rate $\dot{\omega}_1(x)$ is governed by the same linear 
second-order differential equation (\ref{eq:gensol}) as in Section~\ref{sec:elastic-brittle-governing}.
The localized solutions satisfying the boundary conditions have again the form given by (\ref{eq:onset_solution}), i.e.,
\beq\label{eq:mj31}
\dot{\omega}_1(x) = -\frac{\dot{\mu}_1}{2}\left(1+\cos\frac{\sqrt{2}x}{\ell_0}\right), \hskip 10mm -\frac{\pi \ell_0}{\sqrt{2}}\le x \le \frac{\pi \ell_0}{\sqrt{2}}
\eeq
provided that the center of the localized damage zone is at the origin.

\eqref{mj31} describes the initial damage rate, just after localization.
Now we search for possible solutions that 
preserve the size of the damage zone and the shape of the damage profile even after a finite increment.
The assumed form of the solution is thus
\beq\label{eq82}
\omega(x) = A(\mu)   \left(1+\cos\frac{\sqrt{2}x}{\ell_0}\right) , \hskip 10mm -\frac{L_d}{2}\le x \le \frac{L_d}{2}  
\eeq
where $L_d=\sqrt{2}\pi\ell_0$ and $A(\mu)$ is a yet unknown function of the loading parameter $\mu$.
Substituting (\ref{eq82}) into \eqref{mj29}, we find that the assumed type of solution exists if $p=0.5$, and then $A(\mu)=(1-\mu)/2$.
Since $\omega(0)=2A(\mu)$ is the maximum value of damage at the center of the damage zone, $\omega_{\max}$,
we get a linear relation between
the load parameter $\mu$ and the maximum damage. Recall that, for the elastic-brittle model, $\mu=\sigma^2/\ft^2$ where
$\ft$ is the tensile strength. Consequently, $\sigma = \ft\sqrt{1-\omega_{\max}}$.

It is also possible to derive an analytical expression for the inelastic part of elongation corresponding to a given stress
level, similar to (\ref{www}):
\bea\nonumber
w &=& 
\frac{\sigma}{E}\left(\int_{-L_d/2}^{L_d/2} \frac{\dx}{1-\omega(x)}  -L_d \right)
=
\\ \nonumber
&=&
\frac{\sigma}{E}\left(\int_{-\pi\ell_0/\sqrt{2}}^{\pi\ell_0/\sqrt{2}} \frac{2\,\dx}{1+\mu-(1-\mu)\cos\frac{\sqrt{2}x}{\ell_0}}-L_d \right)=
\\
&\label{eq:mj60x}
=&
\frac{\sigma}{E}\left(\frac{2\ell_0}{\sqrt{2\mu}}
\left[\arctan\frac{\tan\frac{x}{\sqrt{2}\ell_0}}{\sqrt{\mu}}\right]_{-\pi\ell_0/\sqrt{2}}^{\pi\ell_0/\sqrt{2}}-L_d \right)=
\frac{\sigma}{E}\left(\frac{\sqrt{2}\pi\ell_0}{\sqrt{\mu}}-L_d \right)
\eea
Taking into account that $\mu=\sigma^2/\ft^2$ and $L_d=\sqrt{2}\pi\ell_0$, we finally get
\beq
w =  \frac{\sigma}{E}\left(\frac{\sqrt{2}\pi\ell_0\ft}{\sigma} - \sqrt{2}\pi\ell_0\right) = \frac{\sqrt{2}\pi\ell_0}{E}(\ft-\sigma)
\eeq
This means that the softening part of the load-displacement diagram is in this case linear and the final elongation at complete
failure is 
\beq
w_{f} = \frac{\sqrt{2}\pi\ell_0 \ft}{E} = \sqrt{2}\pi\ell_0\eps_0 = L_d\eps_0
\eeq
Of course, the load-displacement diagram exhibits snapback, because the elongation at peak load is $L\eps_0$ where $L$
is the total length of the bar, while the final elongation is $L_d\eps_0$, where $L_d=\sqrt{2}\pi\ell_0$ is the size of
the localized damage zone, which must be smaller than the bar length, otherwise the localized solution considered here would not
be valid. Nevertheless, the model is regularized because the total energy dissipated by failure is nonzero and equals
$L_d\eps_0A\ft/2$. The energy dissipated per unit sectional area is thus $\Gf = L_d\eps_0\ft/2 = L_d \gfn$. It corresponds to
the energy dissipated per unit volume, $\gfn$, multiplied by the size of the damage zone, $L_d$,
which is proportional to the
internal length parameter $\ell_0$. 

\begin{figure}[h]
\begin{tabular}{cc}
(a) & (b)
\\
\includegraphics{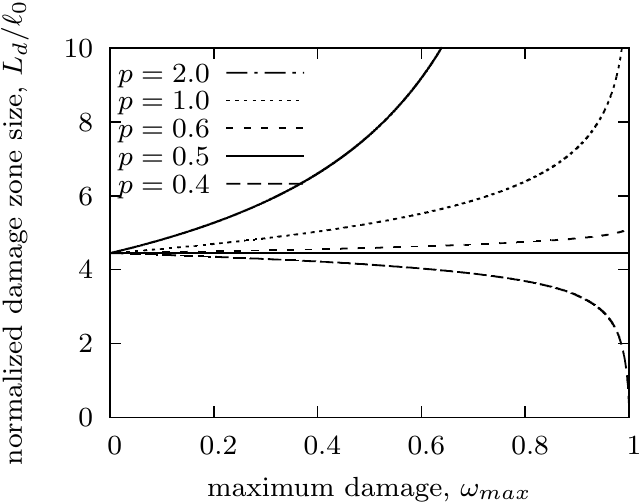}
&
\includegraphics{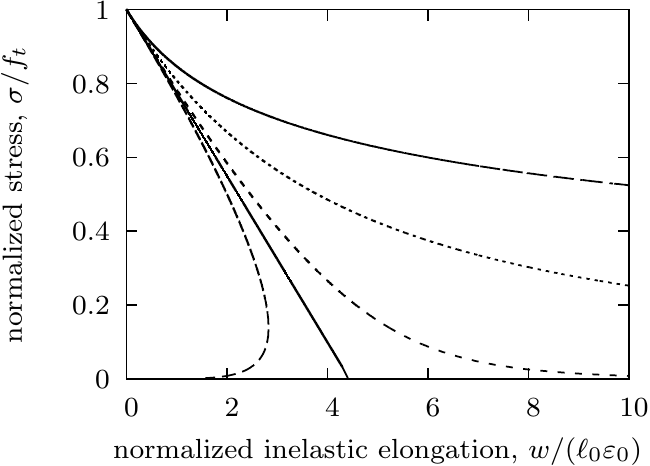}
\\
\includegraphics{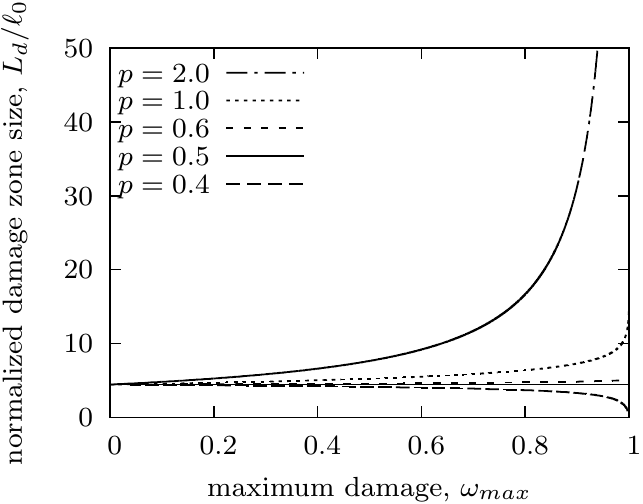}
&
\includegraphics{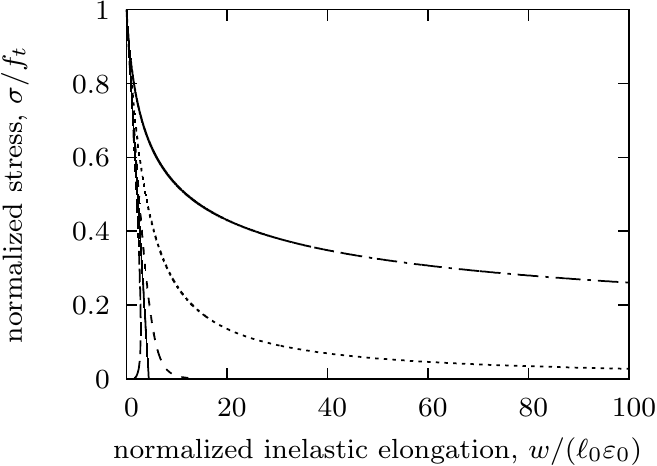}
\end{tabular}
\caption{Regularized elastic-brittle model with variable characteristic length given by \eqref{mj25}
and with different values of exponent $p$:  
(a) size of damage zone (active part) as a function of maximum damage, (b) relation between stress and inelastic part of elongation.}
\label{fig:m5}
\end{figure}

The results of our analysis are confirmed by numerical simulations. \figref{m5}a shows the size of the active part of damage zone as a function
of the maximum damage. 
Note that the top and bottom plots differ only by the scale on the vertical axis.
The horizontal solid line corresponds to $p=0.5$, for which the damage zone remains constant and its
size is $\sqrt{2}\pi\ell_0$, as expected.
For comparison, curves that correspond to selected values of $p$ different from 0.5
are plotted as well. For $p=0.4$, the active part of the damage zone
gradually shrinks to a single cross section. For $p=0.6$, the damage zone expands, but only slightly. The expansion is more
pronounced for higher values of $p$. 
\figref{m5}b shows the inelastic part of the stress-displacement diagram, again for different values of exponent $p$.
It is confirmed that the global softening diagram is linear for $p=0.5$. For lower values of exponent $p$,
the diagram snaps back to the origin, and for higher values it exhibits a tail, which is very long for $p=1$ or 2;
see the bottom plot in~\figref{m5}b.

\begin{figure}[p]
\vspace*{-10mm}
\begin{tabular}{cc}
(a) & (b)
\\
\includegraphics{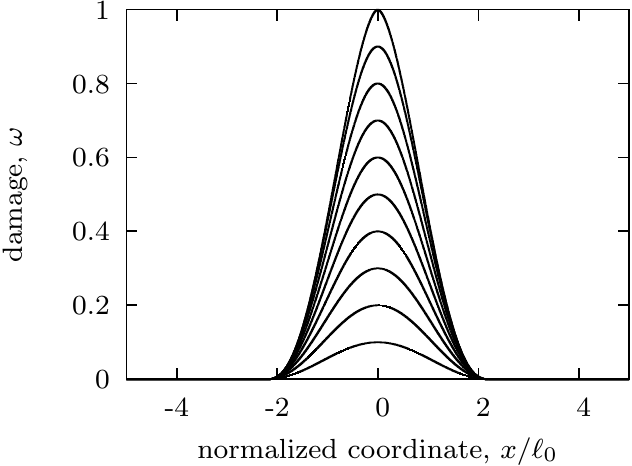}
&
\includegraphics{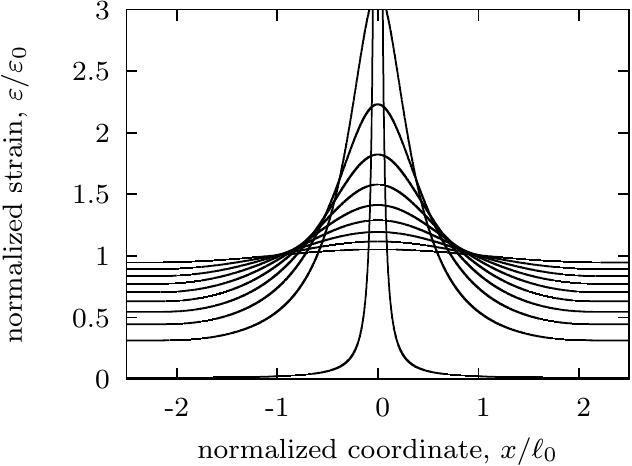}
\\
\includegraphics{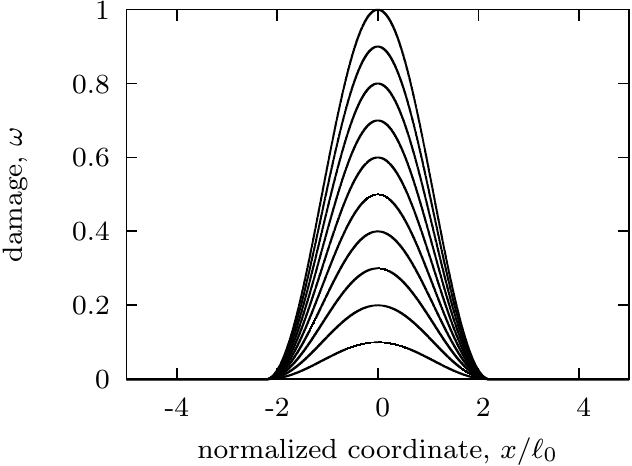}
&
\includegraphics{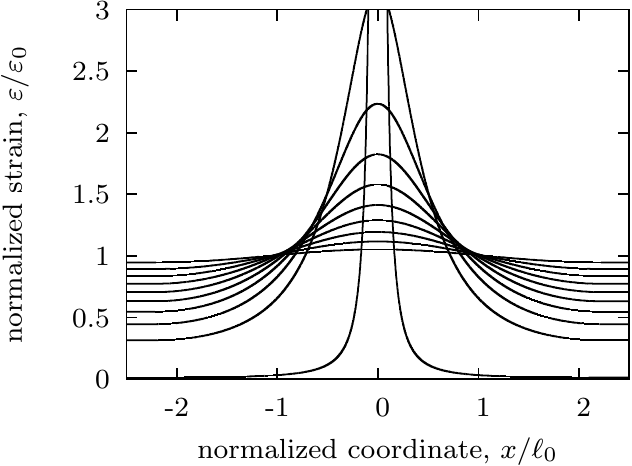}
\\
\includegraphics{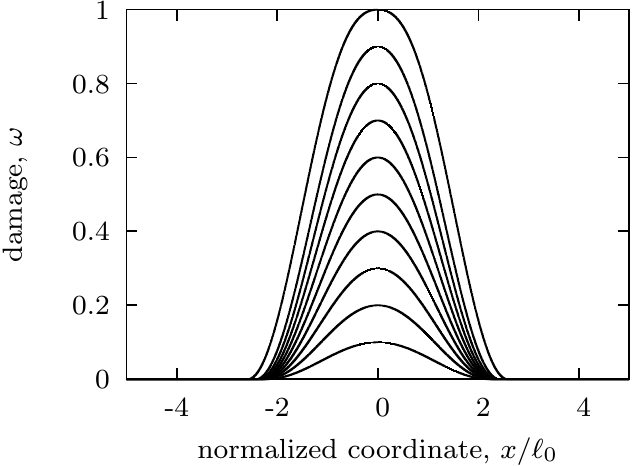}
&
\includegraphics{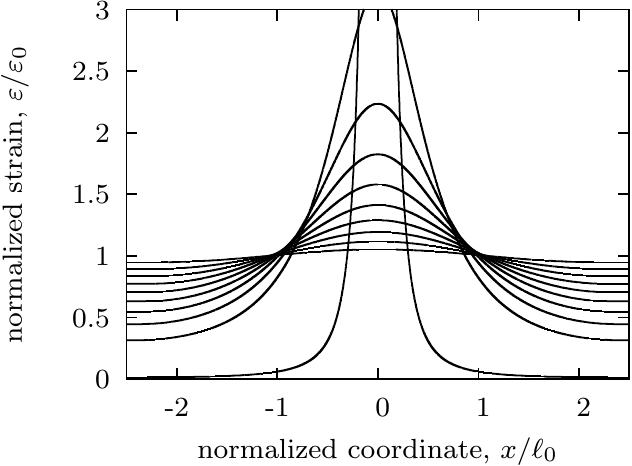}
\\
\includegraphics{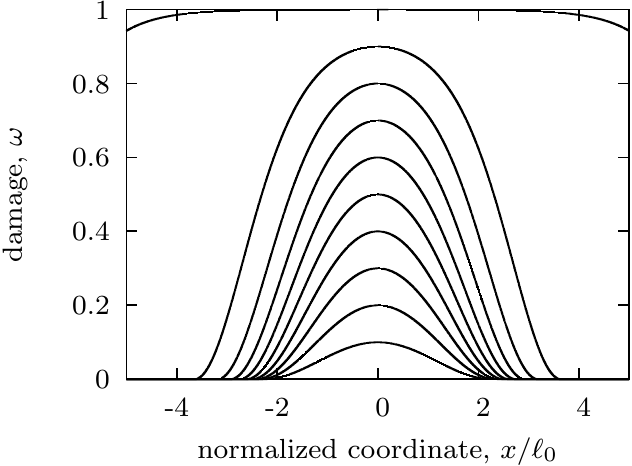}
&
\includegraphics{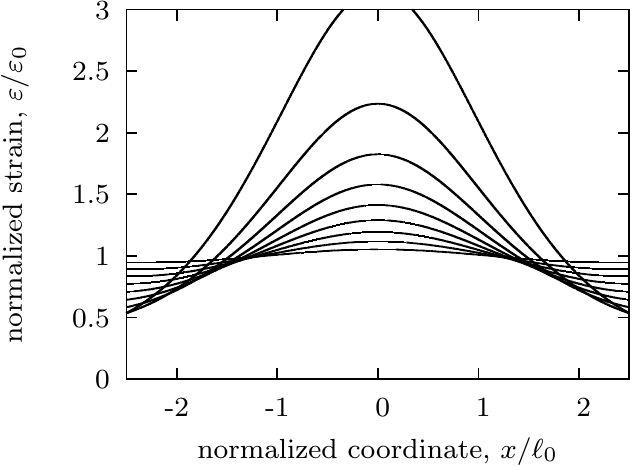}
\end{tabular}
\caption{Regularized elastic-brittle model with variable characteristic length given by \eqref{mj25} and different values
of exponent $p=0.4$, 0.5, 0.6 and 1.0 (from top to bottom):  
evolution of (a)~damage profile, (b)~strain profile.}
\label{fig:m6}
\end{figure}

\begin{figure}[p]
\begin{tabular}{cc}
(a) & (b)
\\
\includegraphics{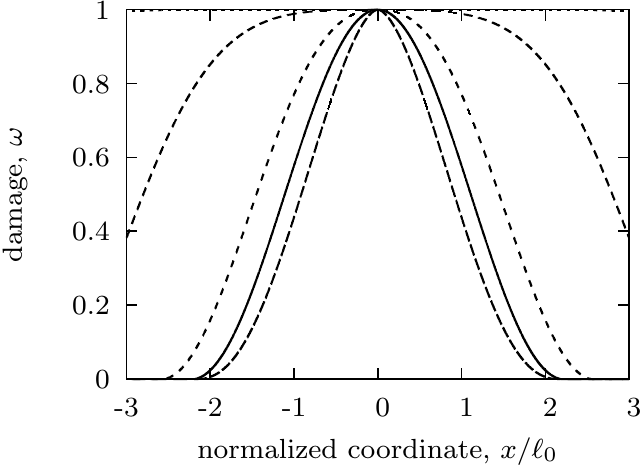}
&
\includegraphics{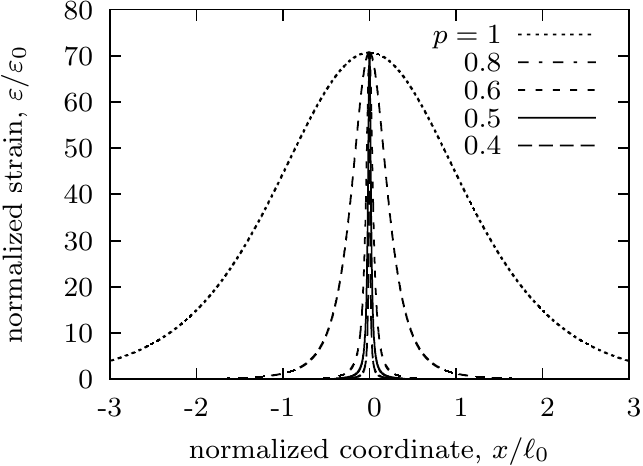}
\\
\includegraphics{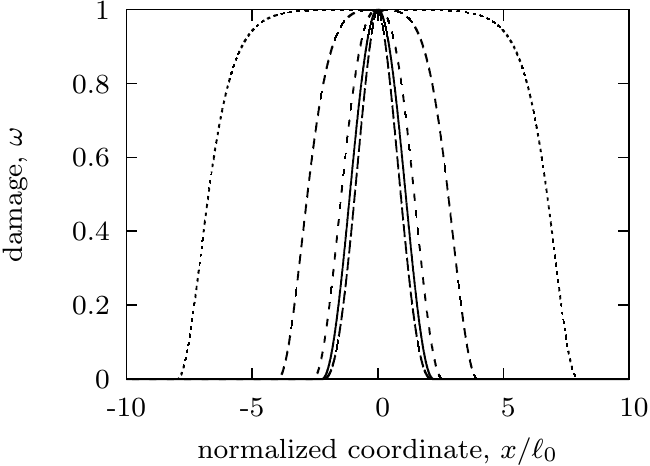}
&
\includegraphics{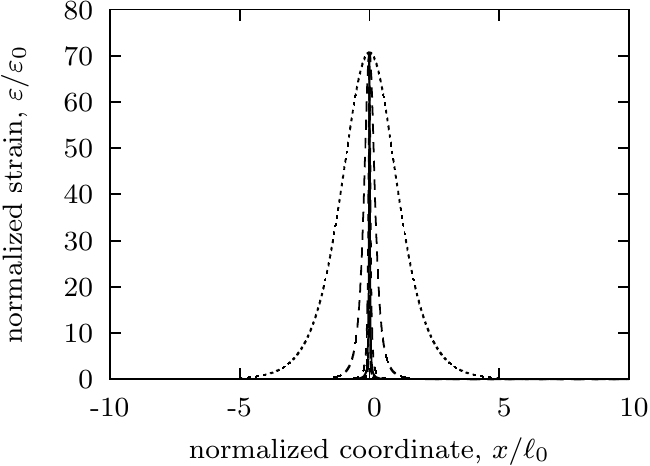}
\end{tabular}
\caption{Regularized elastic-brittle model with variable characteristic length given by \eqref{mj25}:  
comparison of (a) damage profiles and (b) strain profiles at $\omega_{\max}=0.9999$ for different values of exponent $p=1$, 0.8, 0.6, 0.5 and 0.4.}
\label{fig:m9}
\end{figure}

The evolution of damage and strain profiles for $p=0.4$, 0.5, 0.6 and 1.0 (from top to bottom) is plotted in \figref{m6}.
The profiles correspond to values of maximum damage ranging from 0.1 to 0.9 with step 0.1, and the last profile corresponds 
to $\omega_{\max}=0.9999$. The damage zone shrinks for $p<0.5$ and expands for $p>0.5$,
see \figref{m6}a, while the strain profile gets more
concentrated not only for $p<0.5$ but also for somewhat higher values, e.g.\ for $p=0.6$ (but not for $p=1$),
see   \figref{m6}b.
The shapes of the ``final'' profiles of strain and damage 
at $\omega_{\max}=0.9999$ are compared in \figref{m9},
with $p=0.8$ included in the comparison.

\subsection{Alternative interpretation -- approach based on generalized compliance variable}
\begin{figure}[p]
\centering
\includegraphics{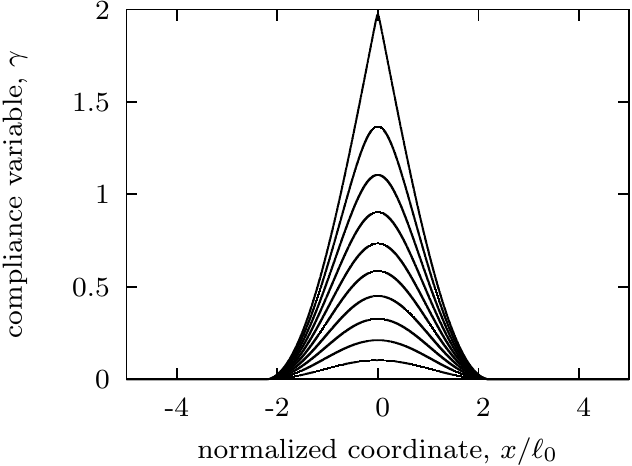}
\caption{Regularized elastic-brittle model with variable characteristic length given by \eqref{mj25} with $p=0.5$:  
profiles of generalized inelastic compliance variable $\gamma$ defined by \eqref{mj70}.}
\label{fig:m10}
\end{figure}

Recall that the modified model with variable characteristic length given by \eqref{mj25} has been motivated by
the special case with $p=2$, which is equivalent to a model with constant characteristic length and gradient
of the inelastic compliance variable $\gamma$ defined in \eqref{mj21}. Even for values of $p$ different from 2,
the regularizing term can be written in the form 
\beq\label{eq:mj22x}
{\cal E}_{\rm reg} = \half\int_{\Omega} \gfn\il \gamma'^2(x)\,{\rm d}x
\eeq
however, with a generalized definition of the inelastic
compliance variable. The appropriate expression for $\gamma$ in terms of $\omega$ can be constructed from
the condition $\ell_0\gamma' = \ell(\omega) \omega'$. This condition can be rewritten as
\beq
\ell_0 \frac{{\rm d}\gamma}{{\rm d}\omega} = \ell(\omega)
\eeq
from which
\beq
\gamma(\omega) - \gamma(0) = \frac{1}{\ell_0}\int_0^{\omega} \ell(\widetilde\omega)\,{\rm d}\widetilde\omega   
\eeq
For the specific dependence of characteristic length on damage given by \eqref{mj25},
evaluation of the integral and application of the condition $\gamma(0)=0$ leads to
\beq\label{eq:mj70bis}
\gamma(\omega)  = \int_0^{\omega} \frac{{\rm d}\widetilde\omega}{(1-\omega)^p} =
\left\{\begin{array}{ll}
 \displaystyle\frac{1-(1-\omega)^{1-p}}{1-p}  & \mbox{ for } p\ne 1
\\
\ln\displaystyle\frac{1}{1-\omega} & \mbox{ for } p= 1
\end{array}\right.
\eeq 
Indeed, for $p=2$ this reduces to the standard definition of inelastic compliance variable (\ref{eq:mj21})
and for $p=0$ to $\gamma(\omega)=\omega$. The special case of $p=0.5$ gives
\beq\label{eq:mj70}
\gamma(\omega)  =  2\left(1-\sqrt{1-\omega}\right)  
\eeq 
So the model that leads to a stationary size of damage zone can be considered as a model with
regularizing term in the form of \eqref{mj22x} in which the inelastic compliance variable is given by \eqref{mj70}.
The evolution of the profiles of this inelastic compliance variable is shown in \figref{m10}.

\section{Modified regularization techniques for  models with softening}
\label{sec6}

In Section~\ref{sec4} it was shown that incorporation of softening (gradual damage growth)
in combination with the simplest regularization approach based on gradient of damage and constant
characteristic length is not sufficient to produce global load-displacement diagrams with a tail,
even if the softening law is constructed such that the local response (in the absence of localization)
is very ductile. Therefore, it is interesting to examine the combination of softening models
with the modified regularization techniques developed in Section~\ref{sec5} for the elastic-brittle model.
This combination provides the most general and most flexible framework, which covers the models from all the
previous sections  as special cases.

A fully general formulation is obtained if the regularizing part of stored energy is defined 
by formula (\ref{eq:mj23}) and the dissipation distance by formula (\ref{eq:GDDm}). 
For convenience, both expressions
are reproduced here:
\bea\label{eq:mj23x}
{\cal E}_{\rm reg}(\tdmg) &=& 
\half\int_{\Omega} \gfn\,\ell^2(\tdmg(x))\, \tdmg'^2(x)\,{\rm d}x
\\
\label{eq:GDDmx}
\DD( \tdmg_1, \tdmg_2 )
&=&
\left\{ 
\begin{array}{ll}
\displaystyle \int_\dmn \left( \Diss(\tdmg_2(x)) - \Diss(\tdmg_1(x)) \right) \de x
  & \mbox{if } \tdmg_2 \geq \tdmg_1 \mbox{ in } \dmn
  \\ +\infty & \mbox{otherwise}
\end{array}
\right.
\eea
Recall that $\ell$ in \eqref{mj23x} is the damage-dependent characteristic length,
which can be defined e.g.\ by formula (\ref{eq:mj25}), and $\Diss$ in (\ref{eq:GDDmx})
is the dissipation density function, the derivative of which is the damage energy release rate $Y$.
For the model with linear softening, the dependence of $Y$ on damage is given explicitly by
\eqref{dissd1}, while for exponential softening it is described implicitly by equations
(\ref{eq:mj2}) and (\ref{eq:inver}). If $\ell$ is set to a constant, $\ell_0$, the formulation
reduces to the simple
regularization technique based on the damage gradient, treated in 
Sections~\ref{sec:elastic-brittle}--\ref{sec4}. 
If $Y$ is set to a constant, $\gfn$, the underlying damage model
reduces to the elastic-brittle model, treated in 
Sections~\ref{sec:elastic-brittle}, \ref{sec:sec3}, and
\ref{sec5}.

\begin{figure}[t]
\begin{tabular}{cc}
(a) & (b)
\\
\includegraphics{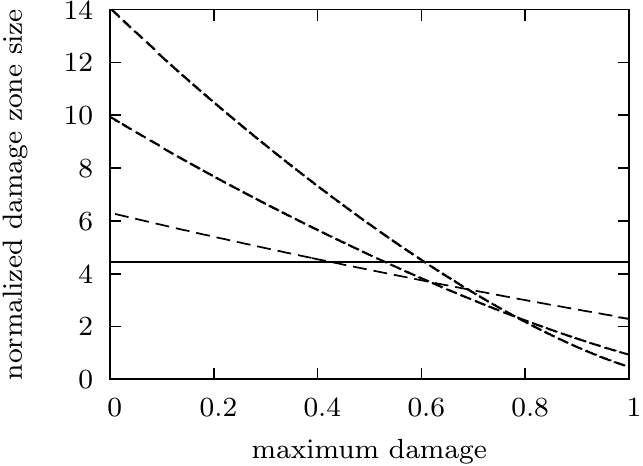}
&
\includegraphics{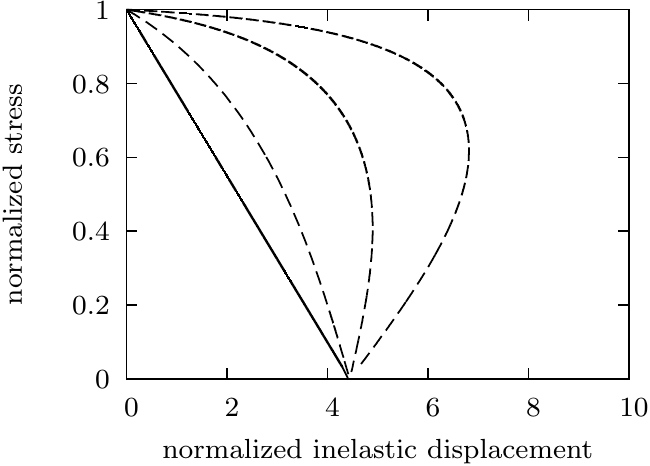}
\end{tabular}
\caption{Regularized {\bf linear softening} model with variable characteristic length given by \eqref{mj25},
with exponent $p=0.5$ and different values of brittleness $\beta=1$, 0.5, 0.2 and 0.1:  
(a) size of damage zone (active part) as a function of maximum damage, (b) relation between stress and inelastic part of elongation.}
\label{fig:m12}
\end{figure}

The differential equation that needs to be satisfied in the damage zone has in fact already
been presented in Section~\ref{sec5.2} as \eqref{mj27}. The analysis in  Section~\ref{sec5.2}
has been performed for the elastic-brittle case only, with the right-hand side of \eqref{mj27}
set to 1. Now we can explore the cases when $Y$ corresponds to linear or exponential softening,
as explained in the previous paragraph. For linear softening with different values of 
brittleness $\beta$ defined in \eqref{beta}, the dependence of the damage zone size on maximum damage
and the inelastic part of the stress-elongation diagram are shown in~\figref{m12}.
For $\beta=1$, the results are plotted by the solid curves
and correspond to the  elastic-brittle model from \secref{sec3}.

The results are also affected by exponent $p$ in formula (\ref{eq:mj25}) for the characteristic length. 
Let us first look at the case of $p=0.5$, for which the elastic-brittle model gives 
a constant size of the damage zone and a linear post-peak part of the stress-elongation diagram.
Numerical results plotted in~\figref{m12}
indicate that with decreasing $\beta$
(i.e., with increasing ductility of the underlying local model), the initial size of the damage zone
increases and the magnitude of the initial post-peak slope decreases, but the final inelastic displacement
at complete failure remains the same, independent of $\beta$. 
This is related to pronounced shrinking of the active part of damage
zone during the localization process. For  lower values
of $\beta$ it shrinks from a higher initial value to a lower final value, which results into an increased
slope of the  stress-elongation diagram and even into snapback. 

\begin{figure}[h]
\begin{tabular}{cc}
(a) & (b)
\\
\includegraphics{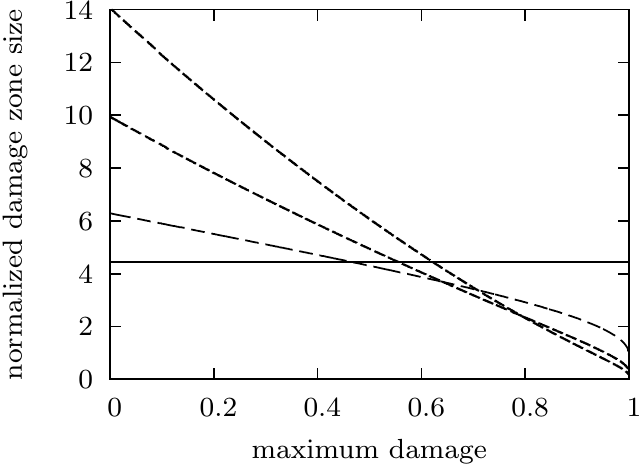}
&
\includegraphics{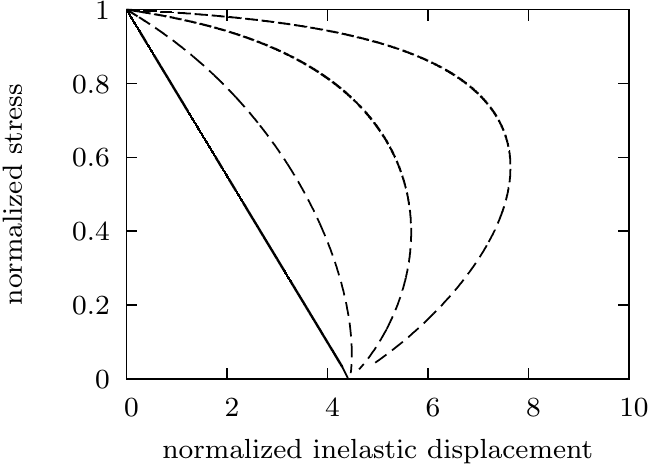}
\end{tabular}
\caption{Regularized {\bf exponential softening} model with variable characteristic length given by \eqref{mj25},
with exponent $p=0.5$ and different values of brittleness $\beta=1$, 0.5, 0.2 and 0.1:  
(a) size of damage zone (active part) as a function of maximum damage, (b) relation between stress and inelastic part of elongation.}
\label{fig:m11}
\end{figure}

Interestingly, the behavior of the model with exponential softening and $p=0.5$ is very similar;
see~\figref{m11}. The final value of inelastic elongation remains
independent of the brittleness number and no tail of the stress-elongation diagram can be produced, even if the local damage
law corresponds to a very ductile behavior (i.e., if $\beta$ is very small). 

Based on the results obtained for the elastic-brittle case, we can expect that a tail should appear
for values of exponent $p$ higher than 0.5, but at the same time there is a danger that the damage zone
would expand.  

\begin{figure}[p]
\begin{tabular}{cc}
(a) & (b)
\\
\includegraphics{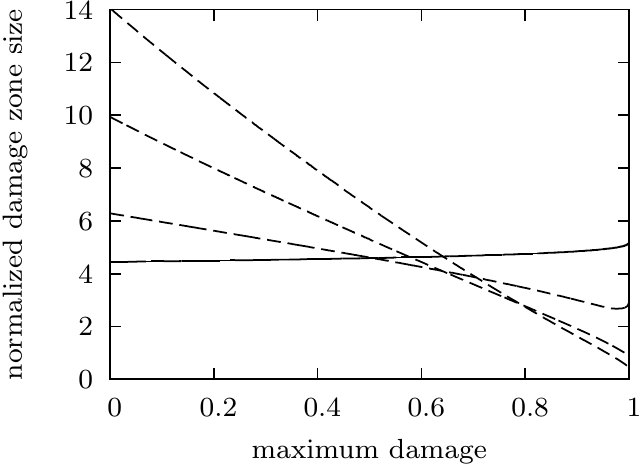}
&
\includegraphics{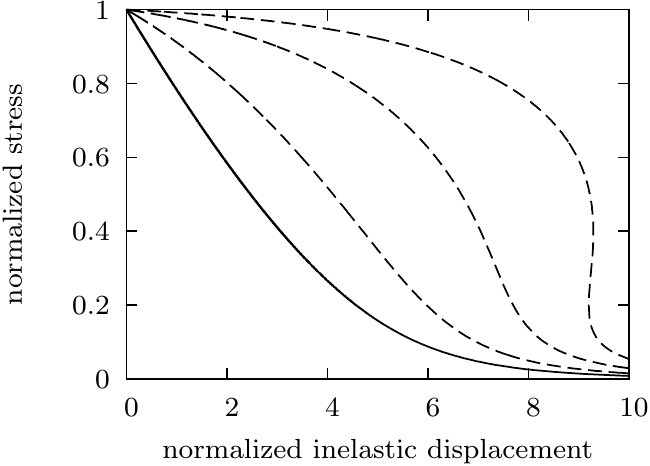}
\\
\includegraphics{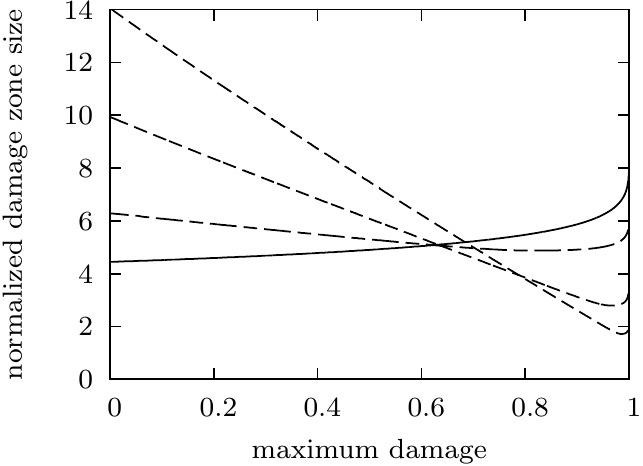}
&
\includegraphics{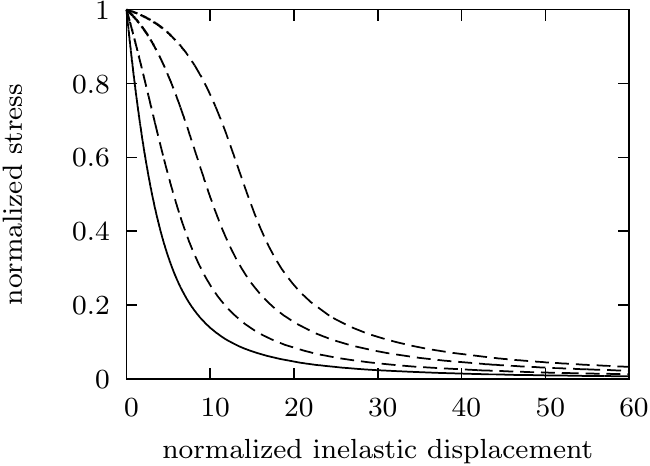}
\\
\includegraphics{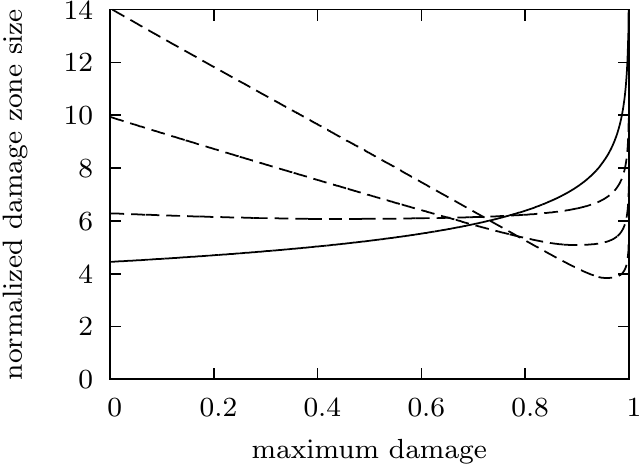}
&
\includegraphics{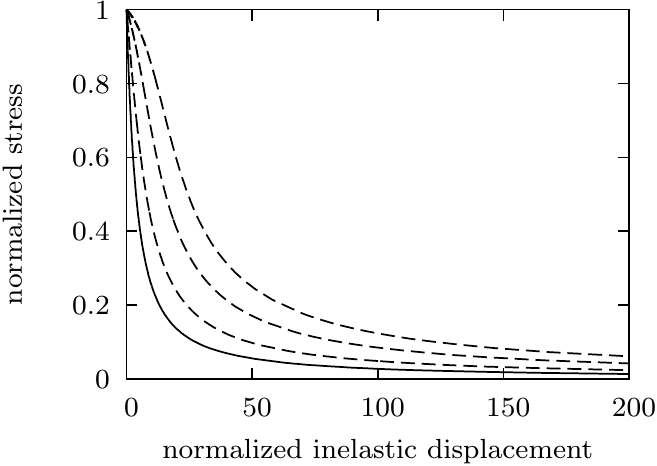}
\end{tabular}
\caption{Regularized {\bf exponential softening} model with variable characteristic length given by \eqref{mj25},
with exponents $p=0.6$, 0.8 and 1.0 (from top to bottom) and different values of brittleness $\beta=1$, 0.5, 0.2 and 0.1:  
(a) size of damage zone (active part) as a function of maximum damage, (b) relation between stress and inelastic part of elongation.}
\label{fig:m14}
\end{figure}

\begin{figure}[p]
\begin{tabular}{cc}
(a) & (b)
\\
\includegraphics{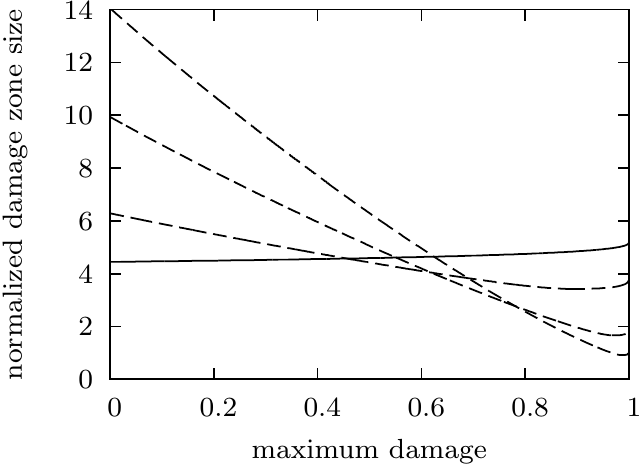}
&
\includegraphics{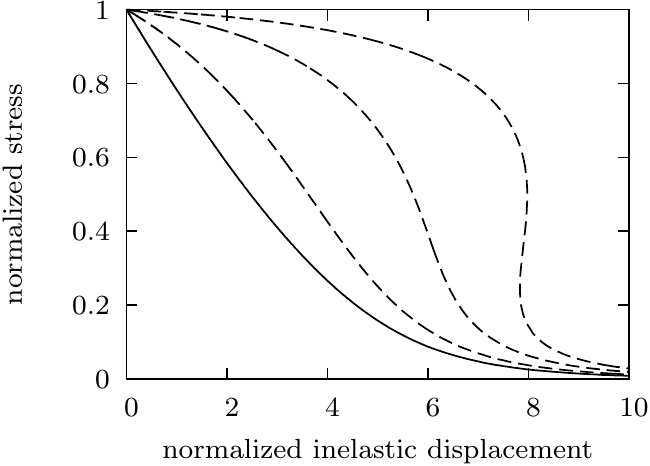}
\\
\includegraphics{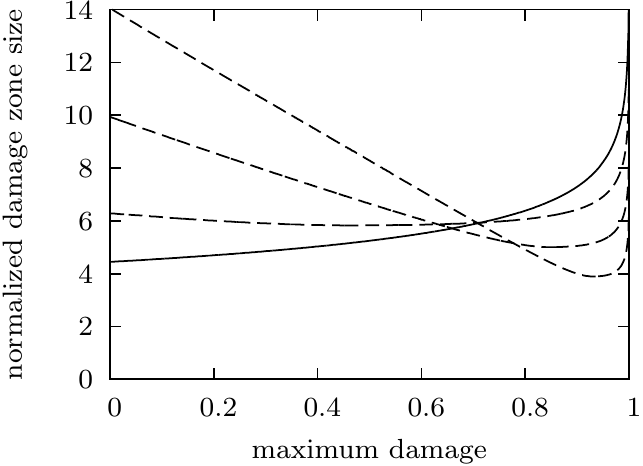}
&
\includegraphics{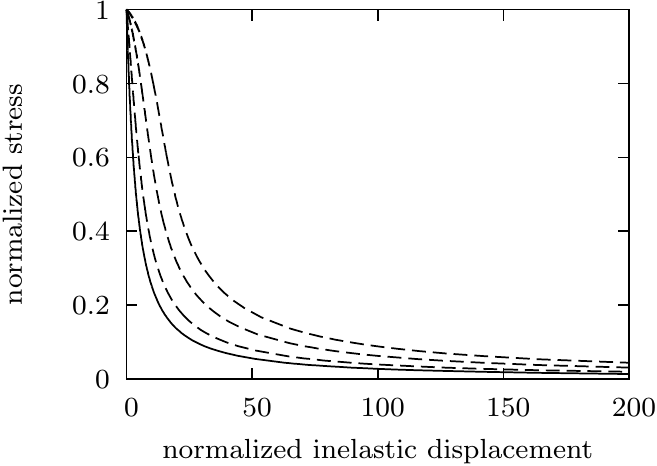}
\end{tabular}
\caption{Regularized {\bf linear softening} model with variable characteristic length given by \eqref{mj25},
with exponents $p=0.6$ and 1.0 (from top to bottom) and different values of brittleness $\beta=1$, 0.5, 0.2 and 0.1:  
(a) size of damage zone (active part) as a function of maximum damage, (b) relation between stress and inelastic part of elongation.}
\label{fig:m15}
\begin{tabular}{cc}
(a) & (b)
\\
\includegraphics{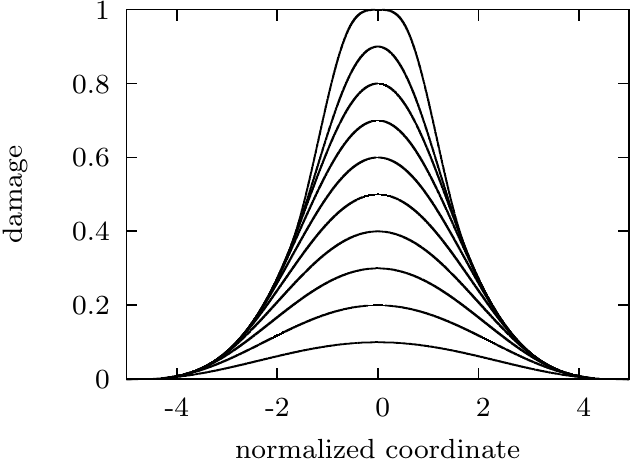}
&
\includegraphics{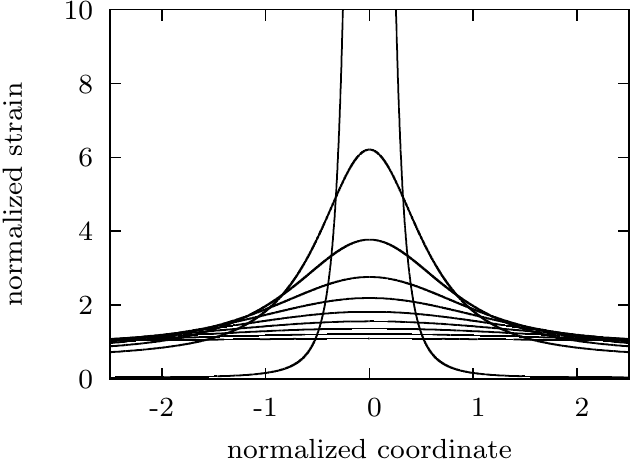}
\end{tabular}
\caption{Regularized {\bf exponential softening} model with variable characteristic length given by \eqref{mj25} and  exponent $p=0.8$ and brittleness $\beta=0.2$:  
evolution of (a)~damage profile, (b)~strain profile.}
\label{fig:m16}
\end{figure}

\section{Summary and conclusions}

Several versions of regularized damage models based on \rev{a
variational} approach have been described and their localization properties have
been examined. All of these models construct the functions describing the displacement field
and the damage field by incremental minimization of a certain energetic functional.
Individual models differ by the specific definitions of stored energy and dissipation distance,
which are needed to construct the corresponding energetic functional. Optimality conditions
lead to a consistent set of equations and inequalities which characterize the displacements and damage 
in the elastic zone,
damage zone, at their interface and at the physical boundary of the body. 

The point of departure was the regularized elastic brittle model with stored energy
consisting of the standard part (\ref{eq:GEstd}) and regularizing part (\ref{eq:GEreg})
and with  dissipation distance (\ref{eq:GDD}). Its localization properties under uniaxial stress were
examined in Section~\ref{sec:sec3} based on (i) linear differential equation (\ref{eq:rate_form})
which characterizes the initial damage rate at bifurcation from a uniform state, and
(ii) nonlinear    differential equation (\ref{eq:dmg_non_lin})
which characterizes the damage profile after a generic incremental step. The linear
rate equation was shown to have an analytical solution while the nonlinear equation was solved
by an incremental-iterative procedure. It was found that the active part of the damage zone
contracts down to zero size during the failure process, and that the dissipated energy is finite
but the load-displacement (or stress-elongation) diagram is quite brittle and exhibits snapback
even for short bars (with respect to the characteristic length of the material).
This kind of model would be suitable for brittle materials but not for quasi-brittle ones,
which typically exhibit a long tail of the load-displacement diagram.

To provide more flexibility and to better control the inelastic part of the  load-displacement diagram,
modified versions of the model were proposed and examined. First of all, softening in the sense
of gradual decrease of stress under uniform conditions was added in Section~\ref{sec4} by generalizing
the dissipation distance; see formulae (\ref{eq:GDDm}) and (\ref{eq:mj2b}). However, numerical simulations revealed
that the enhanced ductility of the material affects only the initial size of the damage zone and the initial
post-peak slope of the  load-displacement diagram. At later stages of the failure process, the damage
zone contracts again to zero size and the   load-displacement diagram exhibits snapback and returns
to the origin. This happens even for a model constructed such that its stress-strain diagram
under uniform strain would have an exponential tail.

As an alternative, a modification of the regularizing part of the stored energy was proposed
in Section~\ref{sec5}. In its simplest form (\ref{eq:mj22}), it replaces the damage gradient by the gradient of the
inelastic compliance variable (\ref{eq:mj21}), which represents a transformed characteristic of damage that tends
to infinity as complete failure is approached (unlike damage, which tends to 1). This simple
modification was found to revert the contraction of the damage zone into its expansion and to produce
an extremely ductile response with stress locking effects.  Therefore, a more general definition
of compliance variable (\ref{eq:mj70bis}) was devised, with parameter $p$ allowing to tune up the
model behavior. As special cases, the original model with regularizing term (\ref{eq:GEreg}) is recovered
for $p=0$ and the simplest form of modified model with regularizing term (\ref{eq:mj22}) is recovered
for $p=2$. It was found that the size of the damage zone remains constant for $p=0.5$, increases for $p>0.5$
and decreases for $p<0.5$. Expansion of the damage zone leads to longer tails of the load-displacement diagram.
It was also demonstrated that the formulation with the gradient of generalized compliance variable is fully
equivalent to a formulation with the gradient of damage combined with a variable characteristic length;
see (\ref{eq:mj23}). 

Finally, a combination of softening incorporated through the dissipation distance with the modified regularization
based on variable characteristic length was examined in Section~\ref{sec6}. It was shown that realistic shapes
of the load-displacement diagram can be obtained for certain combinations
of parameter $p$ and another parameter $\beta$ that describes the ductility under uniform conditions.

\rev{
To keep the present study focused, its scope 
has been limited to localization analysis in the one-dimensional setting,
with physical interpretation in terms of a bar under uniaxial tension. Alternatively,
the same governing equations with a different meaning of individual symbols could describe
an infinite shear layer. Such a simplified analysis has revealed some of the basic features of the
initial model and of its numerous modifications and indicated which general trends could be
expected in a more general setting (e.g., expansion or contraction of the damage zone, and
brittle or ductile character of the load-displacement diagram). Of course, certain other 
aspects cannot be investigated in 1D and would need to be addressed by at least two-dimensional
analytical studies and numerical simulations. 

As pointed out by \cite{Simone:2004:IIP}, simulations of notched specimens
based on integral-type nonlocal damage models exhibit certain pathological or at least questionable
features. For instance, damage is initiated not directly at the tip of a notch or pre-existing crack 
but at a finite distance from the tip, proportional to the characteristic length of the 
nonlocal model. The reason is that integral-type nonlocal models usually consider damage
as driven by nonlocal equivalent strain, obtained by weighted spatial averaging of 
locally evaluated equivalent strains. The local strain has a singularity at the crack tip
but the nonlocal strain remains bounded and turns out to have a maximum at a certain distance
from the tip. It is worth noting that for sufficiently low load levels the maximum nonlocal
strain remains below the damage threshold and the response remains purely elastic, with a stress
singularity. 
In contrast to that, the present gradient-based damage model can be expected
to initiate damage right at the crack tip and for arbitrarily low load levels, because the
singular strain field corresponding to linear elastic fracture mechanics would immediately
violate the multi-dimensional version of condition~(\ref{eq:EL1}), with the
second derivative of damage replaced by the Laplacean of damage and with the term $\half Eu_k'^2$ replaced by the
proper expression for elastic energy density under multiaxial stress.

The issue of damage initiation at or near a notch tip is closely related to the subsequent
evolution of the damage zone and of stresses within this zone, and to the final distribution
of dissipated energy in the vicinity of the notch. As shown in detailed studies of such phenomena
\citep{Jirasek2004,Grassl2014}, 
integral-type nonlocal damage models with standard normalization of the nonlocal weight function
near boundaries lead to excessive values of stresses and dissipation density in the vicinity of
a notch. Such spurious phenomena can be to a large extent reduced by special modifications,
e.g.\ by a reduction of the characteristic length near boundaries. 
Whether the present gradient-based formulation has a similar effect needs to be assessed by 
numerical simulations, which are left for further research.  
}  

\subsection*{Acknowledgments} 
The authors would like to acknowledge the financial support from the Czech
Science Foundation (project No.~108/11/1243). \rev{They would also like to thank
the reviewers for careful reading of the manuscript and for helpful and
stimulating comments.}

\appendix

\section{Numerical simulation of damage evolution}\label{app:algorithm}

Evolution of the localized damage profile can be investigated numerically
by solving a non-linear second-order differential equation that is valid within
the damage zone, with appropriate continuity
conditions imposed at the boundary of this zone. A particular feature of this
problem is that the damage zone evolves and the position of its boundary is not known in advance. 
Numerical treatment of such a problem can be based on a modification of the shooting method.

Let us present the numerical procedure for the  regularized elastic-brittle
model constructed in Section~\ref{sec:elastic-brittle} and analyzed in Section~\ref{sec:sec3}. 
For this model, the governing differential equation (\ref{eq:dmg_non_lin})
reads  
\begin{equation}\label{eq:dmg_non_linx}
\il \dmg_k''(x) + \frac{\mu_k}{( 1 - \dmg_k(x) )^2} = 1
\end{equation}
with $\mu_k = {\sigma_k^2}/{2 E \gf}$ denoting a parameter that equals 1 at the onset of damage and then 
decreases to 0 during the failure process.  
At the onset of damage, we have $\dmg_1(x)=0$ and $\mu_1=1$.
\eqref{dmg_non_linx} should be solved successively for $k=2,3,\ldots N$ 
on intervals $\Omega_{d,k}=(-L_{d,k}/2,L_{d,k}/2)$ of unknown sizes $L_{d,k}$,
taking into account boundary conditions  (\ref{cond16}) and (\ref{cond18}).

Since the damage zone is centered at the origin and the solution is expected to be symmetric
(given by an even function), we can impose symmetry condition (vanishing $\dmg'$) at $x=0$
and compute the values of damage for non-negative $x$ only. 
The size of each step could be controlled by prescribing a certain (negative) increment of parameter $\mu$.
However, it is more convenient to specify the step size by prescribing the value of damage at $x=0$, denoted
as $\dmg_{\max}$, and
consider the value of parameter $\mu$ as unknown. Thus, in a typical step number $k$, we have to find the
values of $\mu_k$ and $L_{d,k}$ such that the solution $\dmg_k(x)$ of \eqref{dmg_non_linx}  satisfies conditions
\bea\label{cond81}
\dmg_k(0) &=& \dmg_{\max,k}\\
\label{cond82}
 \dmg_k'(0) &=& 0 \\
\label{cond83}
\dmg_k(L_{d,k}/2) &=&  \dmg_{k-1}(L_{d,k}/2) \\
\label{cond84}
 \dmg_k'(L_{d,k}/2) &=&\dmg_{k-1}'(L_{d,k}/2)  
\eea

From the numerical point of view,
conditions (\ref{cond81})--(\ref{cond82}) can be considered as initial conditions, which would be sufficient to solve the problem
if $\mu_k$ were known. For a selected trial value of $\mu_k$, we can construct an approximate solution by a finite
difference scheme and find the size of the damage zone $L_{d,k}$ from (\ref{cond84}). If the yet unexploited 
condition (\ref{cond83}) happens to be satisfied, the trial value  of $\mu_k$ is correct and the solution is accepted.
In general, it is necessary to adjust the trial value and iterate on it until  (\ref{cond83}) is satisfied
with a prescribed tolerance. 

To formalize the procedure described above, let us introduce a function
$\tilde\dmg(x,\mu)$ defined as the solution of \eqref{dmg_non_linx} with $\mu_k$ set to $\mu$ and with initial conditions  
(\ref{cond81})--(\ref{cond82}):
\bea\label{a5}
\il \frac{\partial^2\tilde\dmg(x,\mu)}{\partial x^2} + \frac{\mu}{( 1 - \tilde\dmg(x,\mu) )^2} &=& 1
\\
\label{cond91}
\tilde\dmg(0,\mu) &=& \dmg_{\max,k}\\
\label{cond92}
\frac{\partial\tilde\dmg(0,\mu)}{\partial x} &=& 0
\eea
Furthermore, let $\tilde L(\mu)$ be a function defined implicitly as the solution of
\beq\label{a6}
\frac{\partial\tilde\dmg(\tilde L(\mu),\mu)}{\partial x} =  \dmg_{k-1}'(\tilde L(\mu))
\eeq
and let $F(\mu)$ be a function defined as
\beq\label{a7}
F(\mu) = \tilde\dmg(\tilde L(\mu),\mu)-\dmg_{k-1}(\tilde L(\mu))
\eeq
The objective is to find $\mu_k$ such that
\beq
F(\mu_k)=0
\eeq
This can be done iteratively by the Newton method, provided that we are able to evaluate the derivative $\rm{d}F/\rm{d}\mu$. 
Differentiation of  (\ref{a7}) leads to
\beq
\frac{\dd F(\mu)}{\dd\mu} = \frac{\partial\tilde\dmg(\tilde L(\mu),\mu)}{\partial x}\frac{\dd\tilde L(\mu)}{\dd\mu}
+\frac{\partial\tilde\dmg(\tilde L(\mu),\mu)}{\partial\mu}-\frac{\dd\dmg_{k-1}(\tilde L(\mu))}{\dd x}\frac{\dd\tilde L(\mu)}{\dd\mu}
\eeq
By virtue of (\ref{a6}), the first term on the right-hand side cancels with the third term, and the formula
simplifies to
\beq
\frac{\dd F(\mu)}{\dd\mu} = \frac{\partial\tilde\dmg(\tilde L(\mu),\mu)}{\partial\mu}
\eeq

The partial derivative of $\tilde\dmg$ with respect to $\mu$ can be obtained by linearization of 
(\ref{a5})--({cond92}) around the solution $\tilde\dmg(x,\mu)$. 
Differentiation of both sides of (\ref{a5}) with respect to $\mu$ yields
\beq
\il \frac{\partial^3\tilde\dmg(x,\mu)}{\partial x^2\partial\mu} + \frac{1}{( 1 - \tilde\dmg(x,\mu) )^2}+ \frac{2\mu}{( 1 - \tilde\dmg(x,\mu) )^3}\frac{\partial\tilde\dmg(x,\mu)}{\partial\mu} = 0
\eeq
The derivative $\partial\tilde\dmg/\partial\mu$, for convenience denoted as $\tilde\dmg_{\mu}$, can thus be
computed from the linear differential equation
\beq\label{cond90x}
\il \frac{\partial^2\tilde\dmg_{\mu}(x,\mu)}{\partial x^2} + \frac{2\mu}{( 1 - \tilde\dmg(x,\mu) )^3}\tilde\dmg_{\mu}(x,\mu) = - \frac{1}{( 1 - \tilde\dmg(x,\mu) )^2}
\eeq
Since the right-hand sides of (\ref{cond91})--(\ref{cond92}) do not depend on $\mu$, the initial conditions for
$\tilde\dmg_{\mu}$ are
\bea
\label{cond91x}
\tilde\dmg_{\mu}(0,\mu) &=& 0\\
\label{cond92x}
\frac{\partial\tilde\dmg_{\mu}(0,\mu)}{\partial x} &=& 0
\eea

The complete algorithm can be summarized as follows:
\begin{enumerate}
\item Set $k=1$, $\mu_1=1$, $\dmg_1(x)=0$, $\dmg_{\max,1}=0$.
\item Increment the step counter $k$ and set $\dmg_{\max,k}=\dmg_{\max,k-1}+\Delta\dmg_{\max}$.
\item Set $j=0$ and find an initial guess $\mu_k^{(0)}$.
\item By the finite difference method on a grid consisting of points $x_i=i\,\Delta x$, $i=0,1,2,\ldots n$, 
compute an approximate numerical solution $\tilde\dmg_i$
of (\ref{a5})--(\ref{cond92}) with $\mu$ set to $\mu_k^{(j)}$,
and simultaneously compute an approximate numerical solution $\tilde\dmg_{\mu, i}$ of (\ref{cond90x})--(\ref{cond92x}).
Terminate the incrementation of $i$ when $\tilde\dmg_i -\tilde\dmg_{i-1} \le  \dmg_{k-1}(x_i)- \dmg_{k-1}(x_{i-1})$,
and denote the value of counter $i$ for which this happens as $i^*$.
\item Set $F=\tilde\dmg_{i^*} -\dmg_{k-1}(x_{i^*})$ and $F'=\tilde\dmg_{\mu, i^*}$.
\item Increment the iteration counter $j$ and evaluate the updated approximation of parameter $\mu_k^{(j)}=\mu_k^{(j-1)}-F/F'$. 
\item If $\vert F\vert>\epsilon$ (where $\epsilon$ is a prescribed tolerance), go to step 4.
\item Accept the converged solution and define the values 
\beq
\dmg_k(x_i)=\left\{\begin{array}{ll} \tilde\dmg_i & \mbox{ for } 0\le i\le i^* \\
\dmg_{k-1}(x_i) & \mbox{ for } i^*< i\le n \end{array}\right.
\eeq 
\item Print the loading parameter $\mu_k^{(j)}$, damage zone size $L_{d,k}=2x^{i^*}$ and damage values $\dmg_k(x_i)$. 
Compute and print other 
relevant quantities, such as the stress, strains, inelastic elongation and total elongation.
\item If  $\dmg_{\max,k}<1$, go to step 2.
\end{enumerate}
The numerical parameters that need to be specified are the tolerance, $\epsilon$,
the increment of maximum damage, $\Delta\dmg_{\max}$,
the spatial step, $\Delta x$, and the number of spatial steps, $n$, which should be large enough to make
sure that $2n\,\Delta x$ is larger than the maximum possible size of the damage zone. Of course, $\Delta\dmg_{\max}$ and $\Delta x$
could be changed adaptively. It is also necessary to select a finite difference scheme for the discretization in space,
which is applied in step 4. 

For instance, if the central difference scheme is used, Eq.~(\ref{a5}) is replaced by
\beq\label{a20}
\frac{\il}{(\Delta x)^2} \left(\tilde\dmg_{i+1}-2\tilde\dmg_{i}+\tilde\dmg_{i-1}\right)  + \frac{\mu}{( 1 - \tilde\dmg_i )^2} = 1
\eeq
from which
\beq
\tilde\dmg_{i+1}=2\tilde\dmg_{i}-\tilde\dmg_{i-1}+\frac{(\Delta x)^2}{\il}\left(1-\frac{\mu}{( 1 - \tilde\dmg_i )^2}\right)
\eeq
This recursive evaluation is applied for $i=1,2,\ldots i^*$. The values at the first two grid points,
\bea
\tilde\dmg_0 &=& \dmg_{\max,k} \\
\tilde\dmg_1 &=& \tilde\dmg_0 + \frac{(\Delta x)^2}{2\il}\left(1-\frac{\mu}{( 1 - \tilde\dmg_0 )^2}\right)
\eea
 are determined
from the initial conditions (\ref{cond91})--(\ref{cond92}) combined with (\ref{a20}) written for $i=0$.
In a similar fashion, the discretized form of (\ref{cond90x})--(\ref{cond92x}) leads to
\bea
\tilde\dmg_{\mu,0}&=& 0 \\
\tilde\dmg_{\mu,1}&=& -\frac{(\Delta x)^2}{2\il}\frac{1}{( 1 - \tilde\dmg_0 )^2} \\
\tilde\dmg_{\mu,i+1}&=&2\tilde\dmg_{\mu,i}-\tilde\dmg_{\mu,i-1}-\frac{(\Delta x)^2}{\il}\left(\frac{1}{( 1 - \tilde\dmg_i )^2}+\frac{2\mu\tilde\dmg_{\mu,i}}{( 1 - \tilde\dmg_i )^3}\right), \hskip 5mm i=1,2,\ldots i^*\nonumber \\
\eea

In step 3 of the algorithm, an initial guess of the load parameter $\mu_k^{(0)}$ has to be specified.
In the first step (i.g., for $k=2$), the initial guess  $\mu_2^{(0)}=1-\dmg_{\max,2}$
can be obtained from the analytical expression (\ref{eq:onset_solution})
for the initial damage rate.  In the subsequent steps, the initial guess can be computed by extrapolation
of the dependence between $\dmg_{\max}$ and $\mu$  based on their values from the previous steps.

For simplicity, the numerical approach has been presented for \eqref{dmg_non_lin}, which corresponds
to the regularized elastic-brittle model with constant characteristic length. Its extension to \eqref{mj27},
which corresponds to the most general regularized softening model with variable characteristic length and
covers equations (\ref{eq:mj7}), (\ref{eq:mj22x}) or (\ref{eq:mj29}) as special cases,  
is straightforward.

\end{document}